\documentclass[twocolumn]{aastex61}

\received{--}
\revised{--}
\accepted{--}
\submitjournal{ApJ}

\shortauthors{Harada et al.}

\begin{document}

\title{Molecular-Cloud-Scale Chemical Composition III: Constraints of Average Physical Properties through Chemical Models}

\correspondingauthor{Nanase Harada}
\email{harada@asiaa.sinica.edu.tw}

\author[0000-0002-6824-6627]{Nanase Harada}
\affiliation{Academia Sinica Institute of Astronomy and Astrophysics, 11F of AS/NTU Astronomy-Mathematics Building, No.1, Sec. 4, Roosevelt Rd, Taipei 10617, Taiwan}

\author[0000-0003-0563-067X]{Yuri Nishimura}
\affil{Institute of Astronomy, The University of Tokyo, 2-21-1, Osawa, Mitaka, Tokyo 181-0015, Japan}
\affil{Chile Observatory, National Astronomical Observatory of Japan, 2-21-1, Osawa, Mitaka, Tokyo 181-8588, Japan}

\author[0000-0002-9668-3592]{Yoshimasa Watanabe}
\affil{Faculty of Pure and Applied Sciences, University of Tsukuba, 1-1-1, Tennodai, Tsukuba, Ibaraki 305-8577, Japan}
\affil{Tomonaga Center for the History of the Universe, University of Tsukuba, 
1-1-1 Tennodai,Tsukuba, Ibaraki 305-8571, Japan}

\author{Satoshi Yamamoto}
\affil{Department of Physics, The University of Tokyo, 7-3-1, Hongo, Bunkyo-ku, Tokyo 113-0033, Japan}

\author[0000-0003-3283-6884]{Yuri Aikawa}
\affil{Department of Astronomy, The University of Tokyo, 7-3-1, Hongo, Bunkyo-ku, Tokyo 113-0033, Japan}

\author[0000-0002-3297-4497]{Nami Sakai}
\affil{RIKEN, 2-1 Hirosawa, Wako, Saitama 351-0198, Japan}

\author[0000-0002-0095-3624]{Takashi Shimonishi}
\affil{Frontier Research Institute for Interdisciplinary Sciences, Tohoku University, Aramaki- azaaoba 6-3, Aoba-ku, Sendai, Miyagi, 980-8578, Japan}
\affil{Astronomical Institute, Tohoku University, Aramakiazaaoba 6-3, Aoba-ku, Sendai, Miyagi, 980-8578, Japan}




\begin{abstract}
It is important to understand the origin of molecular line intensities and chemical composition in the molecular-cloud scale in the Galactic sources
 because it serves as a benchmark to compare with the chemical compositions of extragalactic sources.
Recent observations of the 3-mm spectra averaged over the 10-pc scale show similar spectral pattern among sources
for molecular lines HCN, HCO$^+$, CCH, HNC, HNCO, c-C$_3$H$_2$, CS, SO, N$_2$H$^+$, and CN.
To constrain the average physical property emitting such spectral pattern,
we model molecular spectra using a time-dependent gas-grain chemical model followed by 
a radiative transfer calculation. We use a grid of physical parameters such as the density $n=3 \times 10^2 - 3\times 10^4$ cm$^{-3}$,
the temperature, $T=10-30$ K, the visual extinction $A_{\rm V} = 2,4,10$ mag, the cosmic-ray ionization rate $\zeta = 10^{-17} - 10^{-16}$ s$^{-1}$, 
and the sulfur elemental abundance $S/H = 8\times 10^{-8} - 8\times 10^{-7}$.
Comparison with the observed spectra indicates that spectra are well reproduced with the relatively low density of $n=(1-3) \times 10^3\,$cm$^{-3}$,
$T=10$\,K, $\zeta = 10^{-17}$ s$^{-1}$, and the short chemistry timescale of $10^5$ yrs. 
This short chemistry timescale may indicate that molecular clouds are constantly affected by the turbulence,
and exposed to low-density, low $A_{\rm V}$ regions that  ``refreshes" the chemical clock by UV radiation.
The relatively low density obtained is orders of magnitude lower than the commonly-quoted critical density in the optically thin case.
Meanwhile, this range of density is consistent with results from recent observational analysis of molecular-cloud-scale mapping.
\end{abstract}




\section{Introduction} \label{sec:intro}
The composition of multiple molecular species sensitively reflects the physical conditions and ages in molecular clouds.
Its evolution along the stages of star formation has been studied in many prestellar- and protostellar cores \citep[e.g., ][]{2009ARA&A..47..427H,2012A&ARv..20...56C,2017iace.book.....Y}.
In addition to an age of a molecular cloud, surrounding environments such as UV radiation, cosmic rays, X-rays, 
or shocks can change the chemical composition 
\citep[see ][ and references therein]{2018IAUS..332...25H}.
Most astrochemical studies on Galactic sources have so far focused on sub-pc regions in the star-forming regions, and we have limited knowledge 
on how the molecular emission of various molecular species distribute in the 10-pc scale molecular cloud. 
Such large-scale observations can reveal various stages of chemical evolution, while the chemistry is more evolved in star-forming regions. 
Recent high-capability interferometers such as Atacama Large Millimeter/sub-millimeter Array (ALMA) have accelerated the development of astrochemical studies in external galaxies.
In those observations, a readily available angular resolution ($\sim 0.5''$) corresponds to the linear scale of 10 pc at a 4-Mpc distance, 
and hence, it is important to understand the chemistry in Galactic molecular clouds with known physical environments in the 10-pc scale.
To resolve this lack of ``benchmark," there have recently been several studies to observe Galactic molecular could complexes in the parsec to the 10-pc scale: W49 in the 1-mm band \citep{2015A&A...577A.127N}, Orion B Molecular Cloud \citep{2017A&A...599A..98P,2017A&A...604A..74S}, Orion A \citep{2017A&A...605L...5K}, 
 W51 \citep{2017ApJ...845..116W}, W3(OH) \citep{2017ApJ...848...17N}, Aquila and Ophiuchus \citep{2017A&A...604A..74S} in the 3-mm band.

Astrochemistry in some external galaxies can be significantly different from the one in the Milky Way,
and the chemistry can be used as probes of their activities.
They may be going through violent stages of active galactic nuclei (AGNs) or starbursts.
Even with single-dish telescopes of $\gtrsim$ 1\,kpc beam, variation of chemical composition has been found between different types of galaxies such as 
starburst galaxies vs. AGN-host galaxies \citep{2015A&A...579A.101A,2018PASJ...70....7N}, luminous infrared galaxies \citep{2011A&A...528A..30C}, and low-metallicity galaxies \citep{2016ApJ...818..161N}.
With ALMA, multi-species observations have been conducted for AGN-containing galaxies \citep{2014PASJ...66...75T,2015PASJ...67....8N,2014A&A...570A..28V,2015A&A...573A.116M}, starburst galaxies \citep{2015ApJ...801...63M}, and a galaxy merger \citep{2017PASJ...69....6U,2018ApJ...855...49H} with the resolution of the 10-pc or 100-pc scale. 
A relatively quiescent spiral arm region in M51 was also studied by \citet{2014ApJ...788....4W}.
Spatial variation of chemical composition can also be studied within individual galaxies in regions of galactic centers, bars, and spiral arms.
To study those galaxies, a benchmark in a quiescent region in the 10-pc scale is of importance.

 Results by the large-scale Galactic astrochemical observations in spiral arm regions by
   \citet{2017ApJ...845..116W} and \citet{2017ApJ...848...17N} as well as observations in an extragalactic spiral-arm region of M51 \citep{2014ApJ...788....4W} indicate that 
 there is very little variation of spectral pattern among different star-forming clouds when averaged in the 10-pc scale, within a factor of a few for most of the detected species. 
 These similar molecular intensity patterns are likely to indicate that the average physical properties in those molecular clouds are also similar.
 This similarity leads us to ask what the physical conditions contributing most to this spectral pattern are. 
 All those regions are located in spiral arms and are free of extreme starburst, X-ray source, or metalicity. 
 With the understanding of such typical physical conditions of relatively quiescent region in the molecular-cloud scale, we are able to 
 highlight the regions with the extreme physical conditions such as higher star formation rate, influence of AGNs, or varied metalicities.

In this paper, we attempt to constrain such physical properties that lead to similar spectra among sources 
by comparing chemical modeling, the spectra of Galactic molecular cloud W51 observed in the 10-pc scale, and the spectra of an extragalactic molecular cloud M51 in the 100-pc scale.
We use a grid of parameters to obtain chemical abundances of molecular clouds using a gas-grain chemical model. 
From the chemical abundances, we also simulate emission intensities by solving radiative transfer, 
and observed spectra are compared with the simulated spectra.
Although grid-model calculations for a wide range of physical parameters have been explored in the extragalactic context 
\citep[e.g.,][]{2009ApJ...696.1466B,2011MNRAS.414.1583B,2017A&A...607A.118V}, we here focus on the detailed analysis
in the smaller physical parameter space that is relevant to molecular-cloud-scale observations of Galactic star-forming regions. 
This paper is organized as follows. Section 2 explains the chemical model and parameters that we use.
Parameters for the radiative transfer code are also discussed in this section.
In Section 3, we present chemical abundances and simulated spectra,
and those results are discussed in Section 4.
Finally, we summarize our results in Section 5.

\section{Model}
\subsection{Chemical Models}
We use a time-dependent, gas-grain model based on $Nautilus$ \citep{2009A&A...493L..49H,2010A&A...522A..42S}.
This is a two-phase model of the gas phase and the ice phase, without the distinction of the ice surface layers and the bulk of the ice.
The reaction network contains 1574 species with 122,496 reactions.
The gas-phase network that we use is based on KIDA\footnote{http://kida.obs.u-bordeaux1.fr} with the implementation of 
deuterated species by \citet{2014isms.confEMF09H} \citep[see also ][]{2014MNRAS.445.1299C,2015A&A...584A.124F} although we do not discuss results for deuterated species.
The grain-surface network refers to the one in \citet{2006A&A...457..927G}.
The gas phase and grain-surface chemistry is connected via adsorption and desorption.
 In addition to the thermal desorption, we include non-thermal desorption by UV-photons \citep[e.g., ][]{2007ApJ...662L..23O} and cosmic rays \citep{1993MNRAS.261...83H}.
 Reactive desorption is also included \citep{2007A&A...467.1103G} with the ratio of the surface-molecule bond-frequency 
 to the frequency at which energy is lost to the grain surface $a=10^{-4}-0.01$ depending on reactions.
We use the elemental abundances commonly used in chemical modeling of dark clouds \citep[``low-metal abundances" in ][]{2008ApJ...680..371W} although we also 
use two enhanced values for sulfur (Table \ref{tab:elem}).
We run our models with 3 different values of sulfur abundances because it is uncertain how much sulfur is depleted on grains
in the 10-pc scale due to less sulfur depletion in photon-dominated regions (PDRs) \citep{2006A&A...456..565G},
and there is variation in the degree of depletion from diffuse to dense clouds \citep{2018arXiv180904978F}.

\begin{deluxetable}{cc}
\tablecaption{A set of elemental abundances. $a(-b)$ stands for $a \times 10^{-b}$\label{tab:elem}}
\tablehead{
\colhead{Element} & \colhead{X/H}
}
\startdata
He & 0.14 \\
N & $2.14(-5)$ \\
O & $1.76(-4)$ \\
C & $7.3(-5)$ \\
S & $8.0(-8)$, $2.5(-7)$, $8.0(-7)$ \\
Si & $8.0(-9)$ \\
Fe & $3.0(-9)$ \\
Na & $2.0(-9)$ \\
Mg & $7.0(-9)$ \\
P &$2.0(-10)$ \\
Cl &$1.0(-9)$ \\
\enddata
\end{deluxetable}

The initial condition for each element is fully molecular for H, atomic for N, O, and He, and ionic for the rest of elements.
We employ so-called a ``pseudo-time-dependent" approach, where physical conditions are kept constant while
the time evolution of the chemistry is calculated by rate equations. 
A more realistic picture is where a molecular cloud evolves from diffuse to dense medium \citep[e.g., ][]{2004ApJ...612..921B,2015A&A...584A.124F}.
Nonetheless, a pseudo-time-dependent approach has historically shown good agreement with observations of dark clouds.
We run chemical models using a grid of physical parameters shown in Table \ref{tab:chem_param}.
We use 5 values of molecular hydrogen density $n_{\rm H2}$, 3 values of the gas temperature, 
3 values of the visual extinction, 2 values of the cosmic-ray ionization rate.
We choose these density and the temperature referring to the values derived by the non-local-thermodynamic-equilibrium analysis
of the multi-line H$_2$CO observations toward the spiral arm region of M51 with the 0.7 - 1 kpc beam (Nishimura et al. in preparation).
In our chemical model, we set the gas and dust temperatures to be equal.
Although these H$_2$CO observations of M51 show the best-fit densities around $3\times10^3 - 3\times 10^4$\,cm$^{-3}$,
we expand our parameter space to $3\times 10^2\,$cm$^{-3}$ following the results of \citet{2017A&A...605L...5K} and \citet{2017A&A...599A..98P}.
Since we consider 3 values for the elemental abundance of sulfur ($S/H$), we run total of 270 models.
Fractional abundances at times $t=10^5$ and $10^6$ yrs are used for the radiative transfer calculation.
In total 540 simulated spectra are produced.

\begin{deluxetable}{ccc}
\tablecaption{Physical parameters used for the chemical model. \label{tab:chem_param}}
\tablehead{
\colhead{Parameter} & \colhead{} & \colhead{Value}
}
\startdata
$n_{\rm H2}$ & Molecular hydrogen density & $3\times 10^2$, $1\times 10^3$, $3\times 10^3$, \\
&& $1\times 10^4$, $3\times 10^4$ cm$^{-3}$ \\
$T_{\rm gas} ({\small=} T_{\rm d})$ &Gas (and Dust) Temperature &10, 20, 30 K\\
$A_{\rm V}$ &Visual extinction & 2, 4, 10 mag\\
$\zeta$ & Cosmic-ray ionization rate & $10^{-17}$, $10^{-16}$ s$^{-1}$\\
\enddata
\tablecomments{The interstellar radiation field $G_{0}=1$ was used for all the models.}
\end{deluxetable}

\subsection{Radiative Transfer}
In order to calculate molecular intensities expected for each set of physical parameter, 
we run a publicly available non-local thermodynamic equilibrium radiative transfer code RADEX \citep{2007A&A...468..627V}.
Input parameters of RADEX are: $N$ -- the column density of the species for which emission intensity is calculated , $\Delta v$ -- the line width , 
$T_{\rm gas}$ -- the gas temperature , $T_{\rm bg}$ -- the background temperature , and $n_{\rm H2}$ -- the H$_{2}$ density . 
We assume a fixed value of $N_{\rm H2}=10^{22}$ cm$^{-2}$ since the chemical model only calculates
 fractional abundances, i.e., abundance ratios of certain species over total hydrogen abundance. 
This column density is not necessarily consistent with the visual extinction because both observations and simulations suggest 
molecular clouds have filamentary and clumpy structures inside \citep{2010A&A...518L.102A,2013ApJ...774L..31I,2015A&A...580A..49I}.
This total column density corresponds to $0.1 - 10$ pc in the density range that we use. 
For the line width, we use the value of 10 km s$^{-1}$ for simplicity. 
This value is somewhat smaller than the line width commonly observed in 10-pc scale molecular clouds.
Although we conduct our analysis with integrated intensities only, the line width may change the optical depth of molecular lines. 
Therefore, we repeated our analysis assuming $N_{\rm H2}=10^{21}$ cm$^{-2}$,
and confirmed that the results do not change significantly for lower optical depths.
We use the CMB temperature for the background temperature $T_{\rm bg} = 2.73\,$K.
Table \ref{tab:lines} lists the transitions that we model for our analysis. 
For CO($J=1-0$) and CI($^3P_1 - ^3P_0$), the modeled intensities are only used to calculate CI/CO intensity ratios.

\begin{deluxetable}{ccc}
\tablecaption{Molecular lines used for the simulated spectra. \label{tab:lines}}
\tablehead{
\colhead{Species} & \colhead{Transition} & \colhead{Frequency (GHz)}
}
\startdata
C$_3$H$_2$ &$2_{1,2}-1_{0,1}$ &85.33889\\
CCH &$N=1-0,J=5/2-3/2$&87.31690\\
CCH  &$N=1-0,J=3/2-3/2$&87.40199\\
HNCO  &$4_{0,4}-3_{0,3}$&87.92524\\
HCN  &$1-0$&88.63160\\
HCO$^+$  &$1-0$&89.18840\\
HNC  &$1-0$&90.66357\\
N$_2$H$^+$ &$1-0$&93.17370\\
 CH$_3$OH  &$2_0-1_0$, A$^+$&96.74137\\
 CS &$2-1$&97.98095\\
SO &$N_J = 2_3-1_2$&99.29987\\
 HNCO  &$5_{0,5}-4_{0,4}$&109.90576\\
 CN  &$N = 1-0, J = 1/2-1/2$&113.16867\\
 CN  &$N = 1-0, J = 3/2-1/2$&113.49492\\
 \hline
 CO &$1-0$ &115.27120 \\
 CI &$^3P_1$ - $^3P_0$&492.16065\\
\enddata
\end{deluxetable}

\section{Results}
\subsection{Fractional abundances}\label{sec:fracabun}
Here we discuss how fractional abundances vary with input physical parameters and the sulfur elemental abundances.
All the fractional abundances that are shown here are in the gas phase.

\subsubsection{Dependence on the density:} 
\begin{figure*}
\gridline{\fig{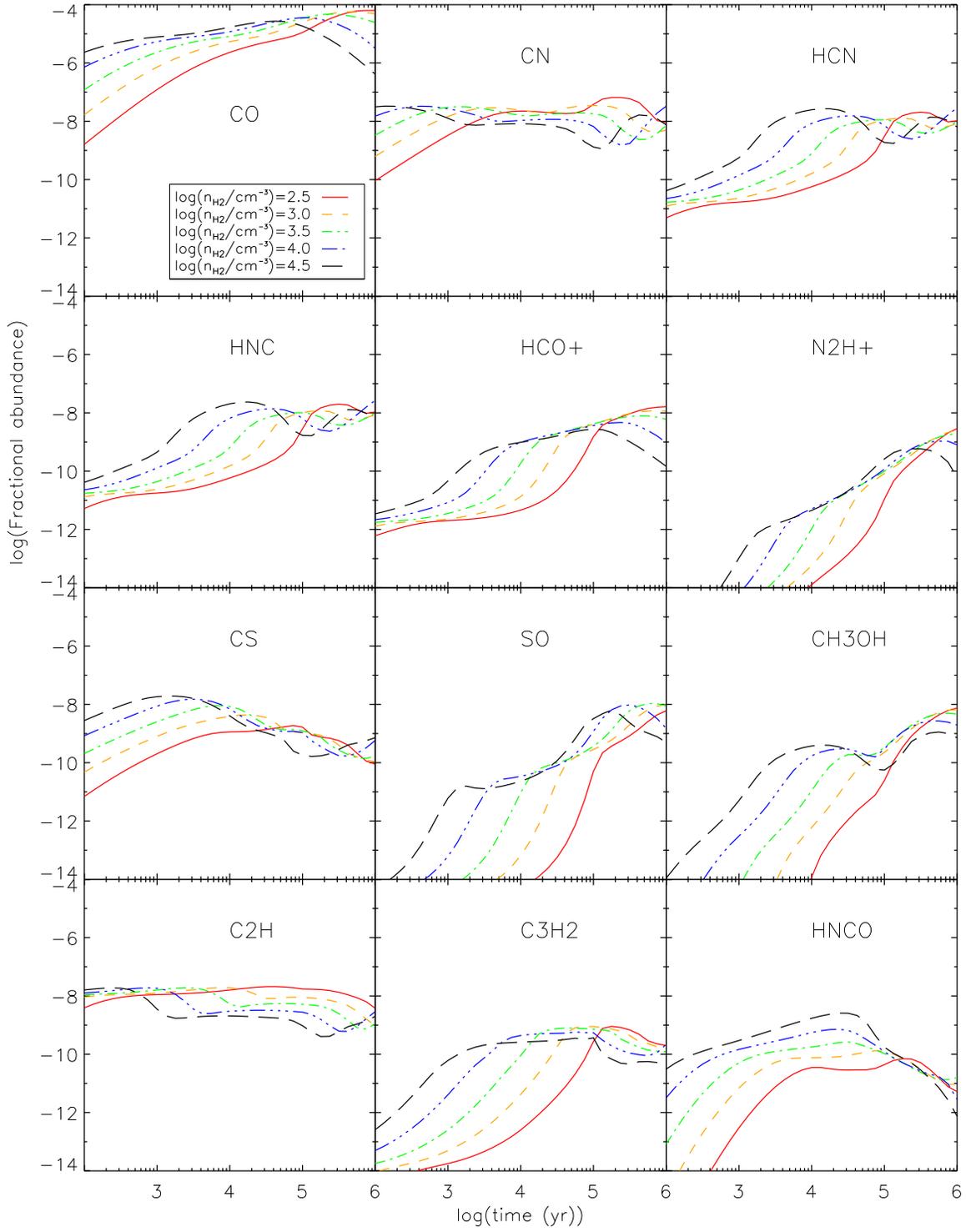}{0.9\textwidth}{}
          }
\caption{Gas-phase fractional abundances of analyzed species with varying $n$ at $T_{gas} = T_{dust} = 10\,$K, 
$\zeta= 1 \times 10^{-17}\,$s$^{-1}$, $A_{\rm V} = 10$ mag, and $S/H = 8 \times 10^{-8}$. \label{fig:n_dep_Av10}}
\end{figure*}

\begin{figure*}
\gridline{\fig{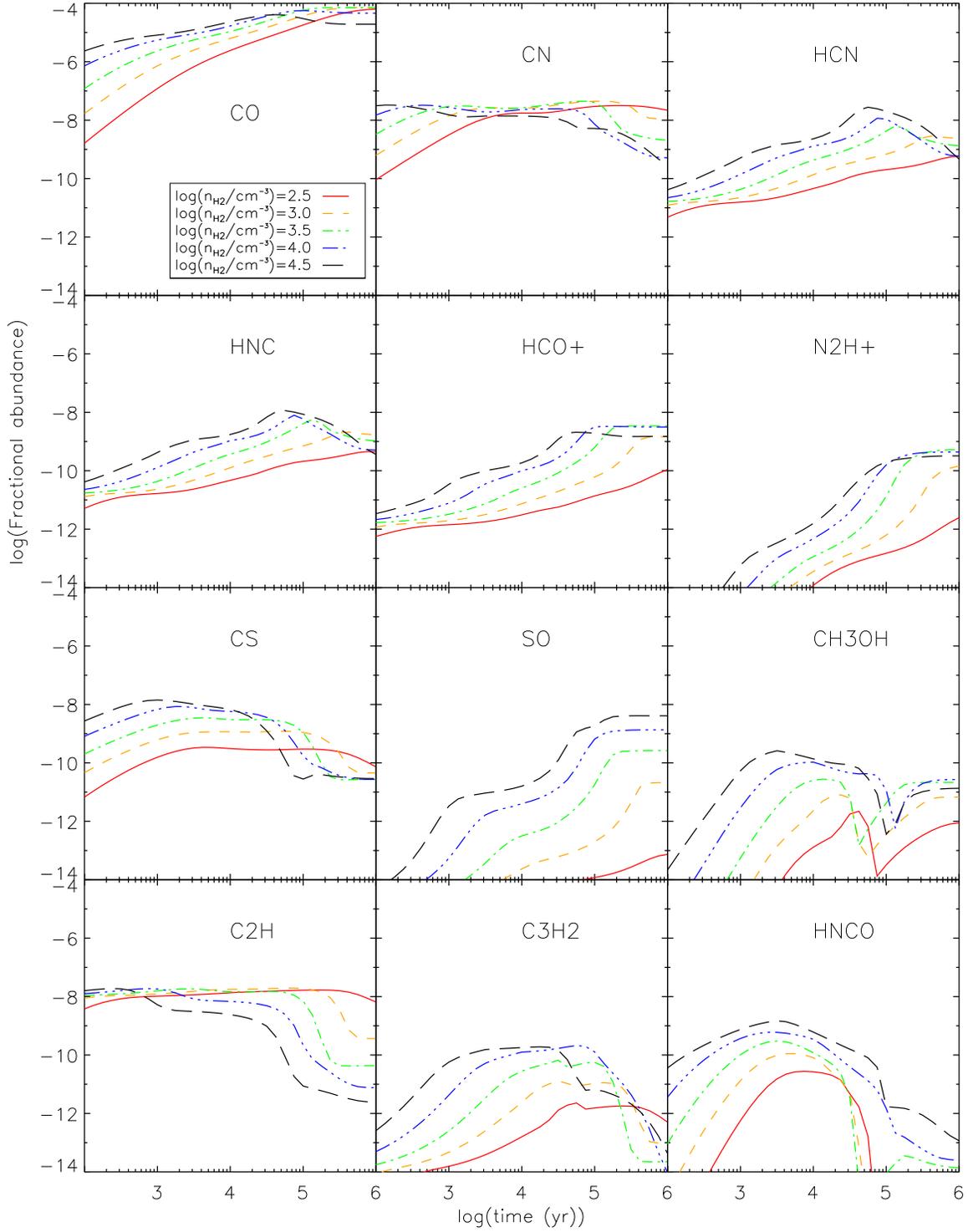}{0.9\textwidth}{}
          }
\caption{Same as \ref{fig:n_dep_Av10}, but with $A_{\rm V} = 2$ mag. \label{fig:n_dep_Av2}}
\end{figure*}

The change in the density mainly affects the chemical time. It becomes shorter with the increasing density.
Figure \ref{fig:n_dep_Av10} shows fractional 
abundances in models with $T= 10\,$K, $\zeta= 1 \times 10^{-17}\,$s$^{-1}$, $A_{\rm V}=10\,$ mag, $S/H = 8 \times 10^{-8}$,
and densities of $n_{\rm H_2}=3\times 10^2\,$cm$^{-3}$, $1\times 10^3\,$cm$^{-3}$, 
$3\times 10^3\,$cm$^{-3}$, $1\times 10^4\,$cm$^{-3}$, and $3\times 10^4\,$cm$^{-3}$.
Figure \ref{fig:n_dep_Av10} shows that the peak abundance does not depend on the gas density, except for HNCO.
Isocyanic acid (HNCO) is produced by a series of surface reactions
\begin{eqnarray}
{\rm CN(s) + O(s) \longrightarrow OCN(s)} \\
{\rm OCN(s) + H(s) \longrightarrow HNCO},
\end{eqnarray}
where ``(s)" indicates species in the ice form, and destroyed by an ion-neutral reaction with He$^+$.
Since the ionization fraction is lower in higher density environments, such reactions with He$^+$ are slower
for the higher density in comparison with other reactions.
At low visual extinction $A_{\rm V}=2$\,mag, photodissociation competes with molecular formation in lower-density condition \footnote{Rates of photodissociation or 
photoionization have the form ${d[A]}/{dt} \propto -G_0 [A]exp(- \gamma A_{\rm V})$ where [X] is the number density 
per unit volume of species X and $\gamma$ is a constant usually having values of 2-3.
On the other hand, rates of 2-body reactions $A + B \longrightarrow C + D$ scale as ${d[C]}/{dt} \propto [A][B]$.
Since the 2-body reactions have terms square of densities while photodissociation or photoionization reactions have term proportional to the density,
photon-related reactions are more prominent at lower densities.
\label{ftn:photodis}}.
Figure \ref{fig:n_dep_Av2} shows fractional abundances as in Figure \ref{fig:n_dep_Av10}
but with $A_{\rm V}=2\,$ mag. For most of the species other than CO, the peak abundances decrease with 
decreasing densities. Exceptions are CCH and CN, whose fractional abundances at the lowest density $n=3\times10^2\,$cm$^{-3}$
are higher than those in cases of higher densities at $t> 1000$ yrs.

\subsubsection{Dependence on the visual extinction} 
\begin{figure*}
\gridline{\fig{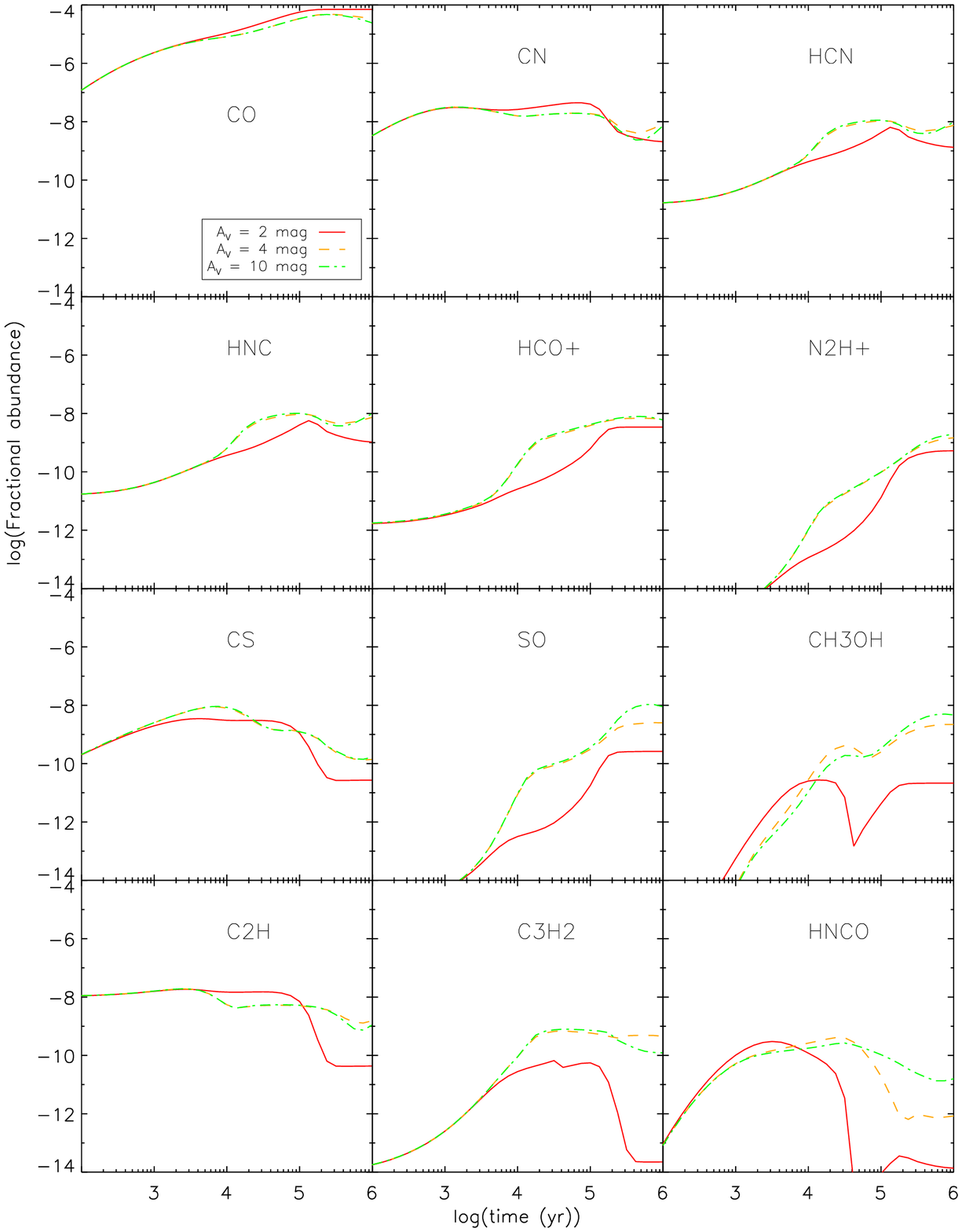}{0.9\textwidth}{}
          }
\caption{Fractional abundances of analyzed species with varying $A_{\rm V}$ at $n_{\rm H2} = 3\times10^3$ cm$^{-3}$, $T_{gas} = T_{dust} = 10\,$K,
$\zeta= 1 \times 10^{-17}\,$s$^{-1}$, and $S/H = 8 \times 10^{-8}$.\label{fig:Av_dep}}
\end{figure*}

\begin{figure*}
\gridline{\fig{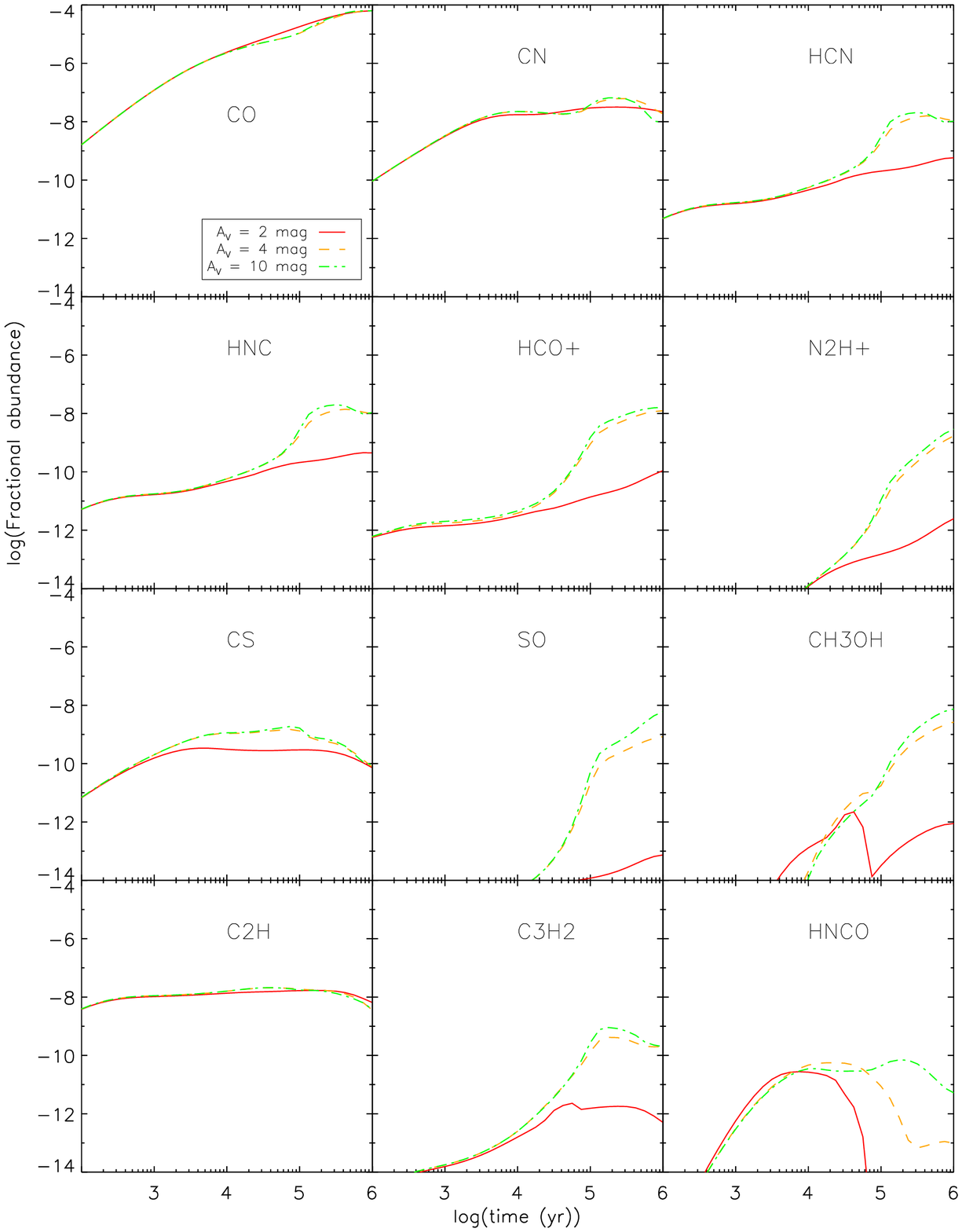}{0.9\textwidth}{}
          }
\caption{Fractional abundances of analyzed species with varying $A_{\rm V}$ at $n_{\rm H2} = 3\times10^2$ cm$^{-3}$, $T_{gas} = T_{dust} = 10\,$K,
$\zeta= 1 \times 10^{-17}\,$s$^{-1}$, and $S/H = 8 \times 10^{-8}$.\label{fig:Av_dep2}}
\end{figure*}

\begin{figure*}
\gridline{\fig{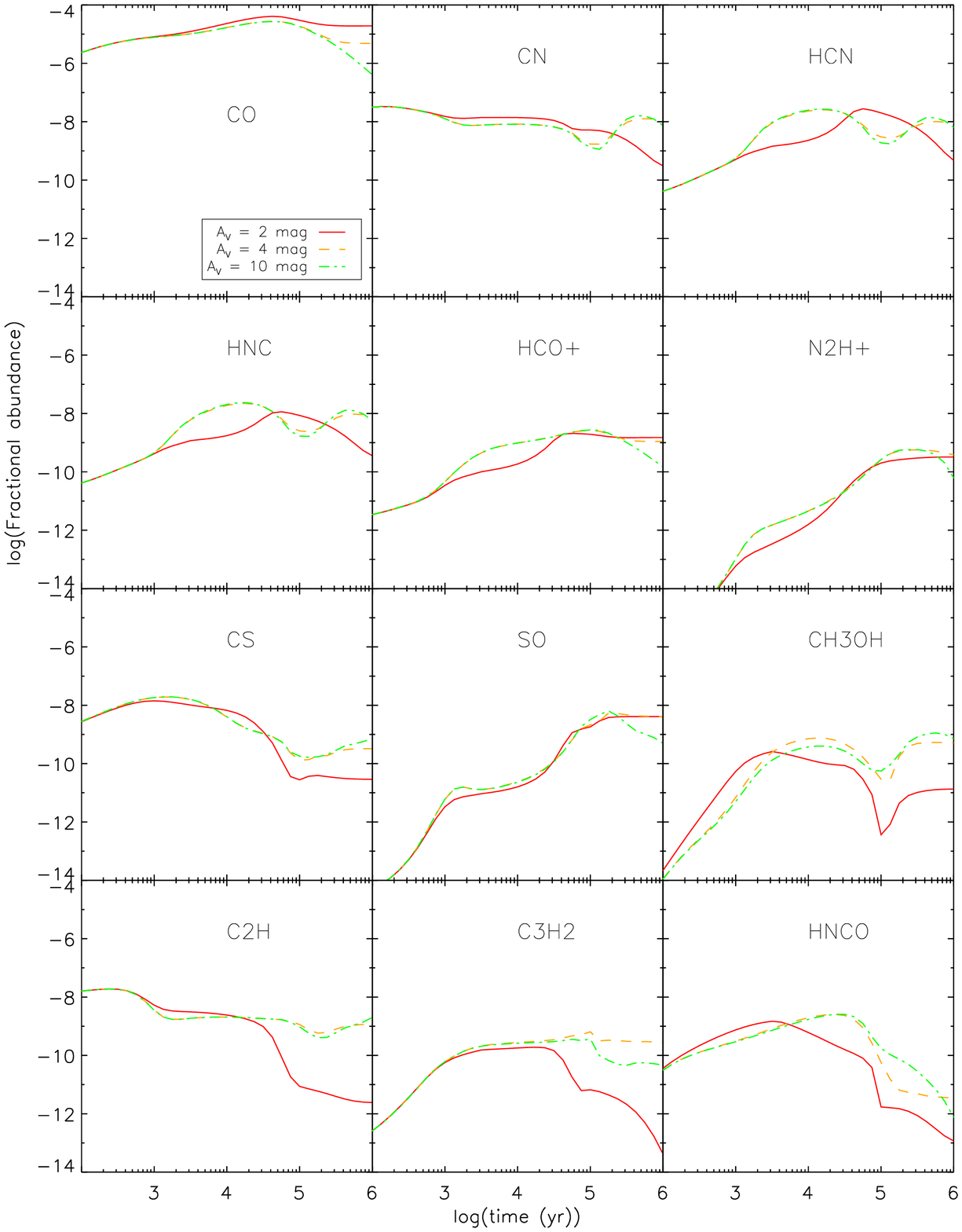}{0.9\textwidth}{}
          }
\caption{Fractional abundances of analyzed species with varying $A_{\rm V}$ at $n_{\rm H2} = 3\times10^4$ cm$^{-3}$, $T_{gas} = T_{dust} = 10\,$K,
$\zeta= 1 \times 10^{-17}\,$s$^{-1}$, and $S/H = 8 \times 10^{-8}$.\label{fig:Av_dep3}}
\end{figure*}

At the low visual extinction, photoionization reactions and photodissociation reactions become effective.
As mentioned earlier, the effect of UV-photons is more prominent for lower densities.
At $n_{\rm H2}=3\times 10^3\,$cm$^{-3}$, fractional abundances of most of the species shown in Figure \ref{fig:Av_dep}
decrease at $A_{\rm V}=2\,$mag except for CO, CN, and CCH. 
The decrease in fractional abundances is most clearly seen for C$_3$H$_2$, CH$_3$OH, and SO, because the UV-photons dissociate these molecules.
For SO, the destruction via the reaction with C$^+$ is also efficient,
and the products of this reaction are ${\rm S^+}$ + CO, S + CO$^+$, and O + CS$^+$ with the same branching ratios.
At a lower density of $n=3\times10^2\,$cm$^{-3}$, the effect of UV-photons is stronger.
Although fractional abundances of CN and CCH are enhanced at $A_{\rm V}=2$\,mag at time $t \sim 10^5$ yr when $n_{\rm H2}=3\times 10^3\,$cm$^{-3}$,
they do not enhance when $n=3\times10^2\,$cm$^{-3}$ (Figure \ref{fig:Av_dep2}).
These effects of UV-photons become less obvious in the case of a higher density $n_{\rm H2}=3\times 10^4\,$cm$^{-3}$, and 
the differences of fractional abundances between the cases of $A_{\rm V}=2\,$mag and $A_{\rm V}=4\,$mag become less compared with the case of $n_{\rm H2}=3\times 10^3\,$cm$^{-3}$
(Figure \ref{fig:Av_dep3}).
For all densities, there are little differences between cases of $A_{\rm V}=4\,$mag and $A_{\rm V}=10\,$mag because 
photodissociation or photoionization reactions are already not effective at $A_{\rm V}=4\,$mag.

\subsubsection{Dependence on the temperature} 
\begin{figure*}
\gridline{\fig{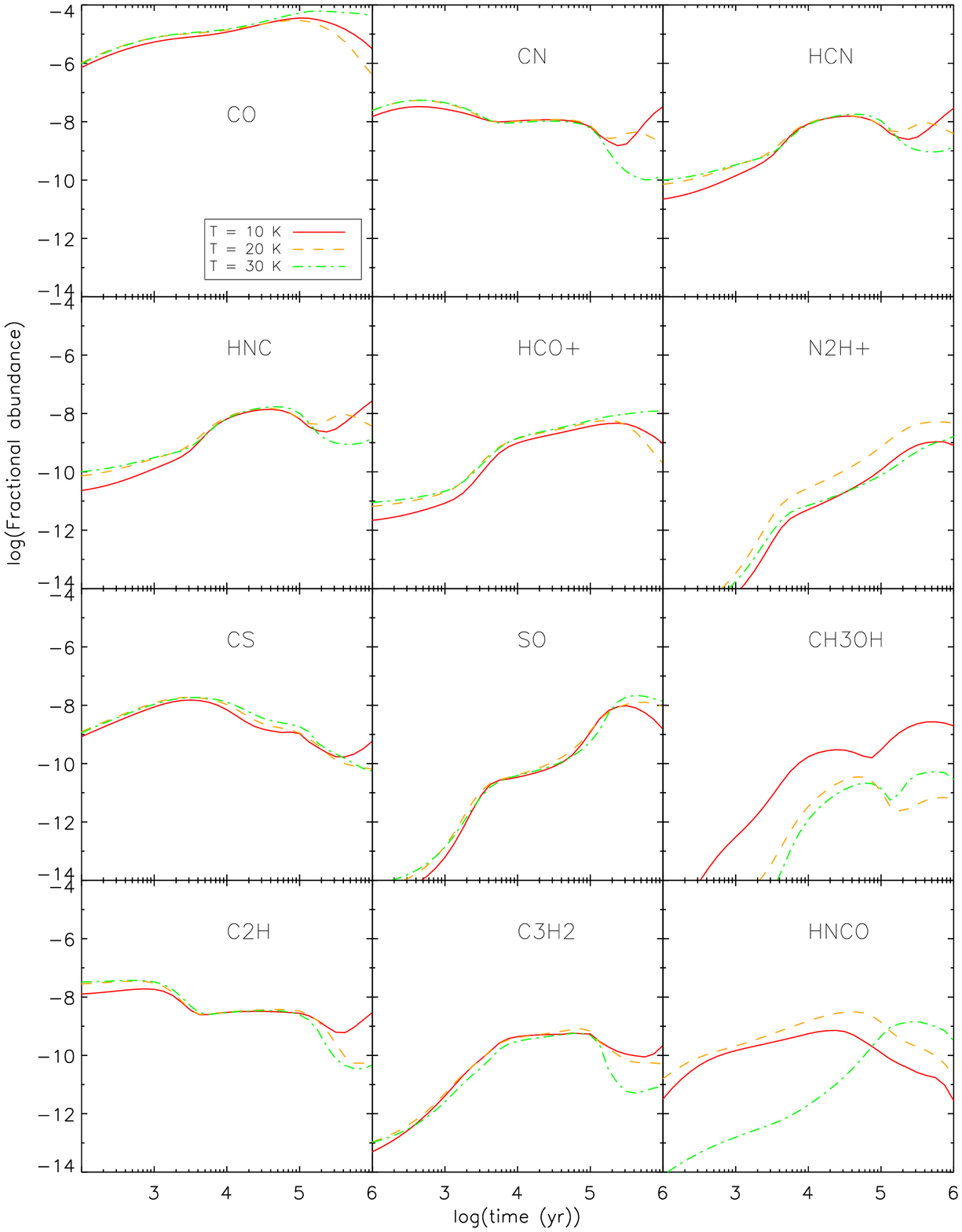}{0.9\textwidth}{}
          }
\caption{Fractional abundances of analyzed species with varying $T$ at $n_{\rm H2} = 10^4$ cm$^{-3}$
$\zeta= 1 \times 10^{-17}\,$s$^{-1}$, $A_{\rm V} = 10$ mag, and $S/H = 8 \times 10^{-8}$. \label{fig:T_dep}}
\end{figure*}

While gas-phase reactions in general do not show strong temperature dependence between 10 - 30 K, 
such a temperature change alters the thermal desorption rate from the ice phase into the gas phase.
Two species with strong dependence on the temperature are CH$_3$OH and HNCO.
Figure \ref{fig:T_dep} shows fractional abundances at $n_{\rm H2} = 1 \times 10^4\,$cm$^{-3}$,
 $\zeta= 1 \times 10^{-17}\,$s$^{-1}$, $A_{\rm V}=10\,$ mag, and $S/H = 8 \times 10^{-8}$.
 Fractional abundances of CH$_3$OH decrease with the increasing temperature; 
 the formation of CH$_3$OH on the grain surface is severely hindered by 
 the fast evaporation of atomic hydrogen at the high temperatures, which leaves very little time for H to react with CO.
 This temperature dependence of CH$_3$OH was already discussed in \citet{2015ApJ...812..142A}.
On the other hand, fractional abundances of HNCO are the highest at $T= 20\,$K.
At $T = 30\,$K, the peak fractional abundance is by a factor of a few higher than that at $T = 10\,$K and is reached at later time.
Although HNCO is also formed on grains, the fractional abundances of HNCO do not decrease at $T= 20- 30\,$K 
because OCN is made efficiently on grains from CN ice at such relatively warm condition. At $T=10$\,K, the main destruction route of CN is through H atom on grains 
to form HCN. However, at $T \gtrsim 20$\,K, the destruction route via H is much slower, and the main destruction route of CN becomes the one through O atom to produce OCN.
Although the H atom on grains becomes less abundant by orders of magnitude at higher temperatures, 
the efficient formation of HNCO on grains is still effective.

\subsubsection{Dependence on the cosmic-ray ionization rate} 
\begin{figure*}
\gridline{\fig{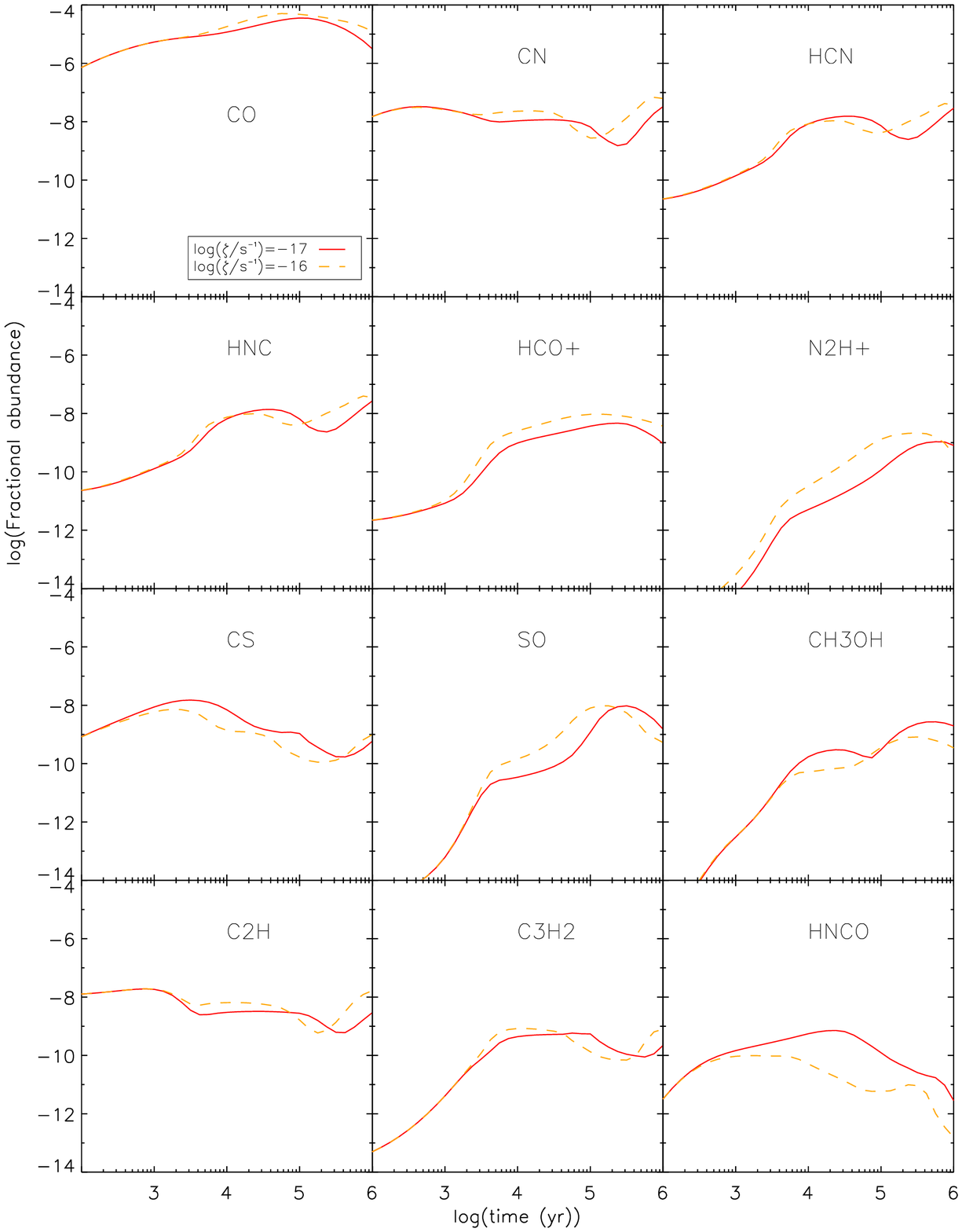}{0.9\textwidth}{}
          }
\caption{Fractional abundances of analyzed species with varying $\zeta$ at $n_{\rm H2} = 1\times10^4$ cm$^{-3}$, $T_{gas} = T_{dust} = 10\,$K,
$A_{\rm V} = 10$ mag., and $S/H = 8 \times 10^{-8}$.\label{fig:z_dep}}
\end{figure*}

Cosmic rays cause ionization reactions of atoms or molecules, and photodissociation of molecules by secondary UV-photons.
Figure \ref{fig:z_dep} shows fractional abundances in models with $n_{\rm H2} = 1\times10^4$ cm$^{-3}$, $T = 10\,$K,
$A_{\rm V} = 10$ mag, and $S/H = 8 \times 10^{-8}$ for $\zeta = 10^{-17}$ and $10^{-16}$ s$^{-1}$.
An order of magnitude enhancement of cosmic-ray ionization rate results in the variation of fractional abundances of most species 
only less than a factor of a few. Species with a relatively strong dependence on the cosmic-ray ionization rate are
N$_2$H$^+$, HCO$^+$, CS, SO, and HNCO.
N$_2$H$^+$ and HCO$^+$ are produced via a reaction
\begin{equation}
{\rm H_3^+ + N_2 \longrightarrow H_2 + N_2H^+,}
\end{equation}
and destroyed via recombination with an electron or a reaction:
\begin{equation}
{\rm N_2H^+ + CO \longrightarrow N_2 + HCO^+.}
\end{equation}
Since the H$_3^+$ fractional abundance increases as a higher cosmic-ray ionization rate, 
N$_2$H$^+$ also increases.
The higher cosmic-ray ionization rate increases the fractional abundances of SO; it is made via a reaction with OH,
while OH is formed via ion-molecule reactions and subsequent dissociative recombination. Thus, SO increases with the higher ionization rate.
Since HNCO is destroyed by He$^+$ or H$^+$, increased ionization fraction decreases the HNCO fractional abundance.

\subsubsection{Dependence on the elemental sulfur abundance} 
\begin{figure*}
\gridline{\fig{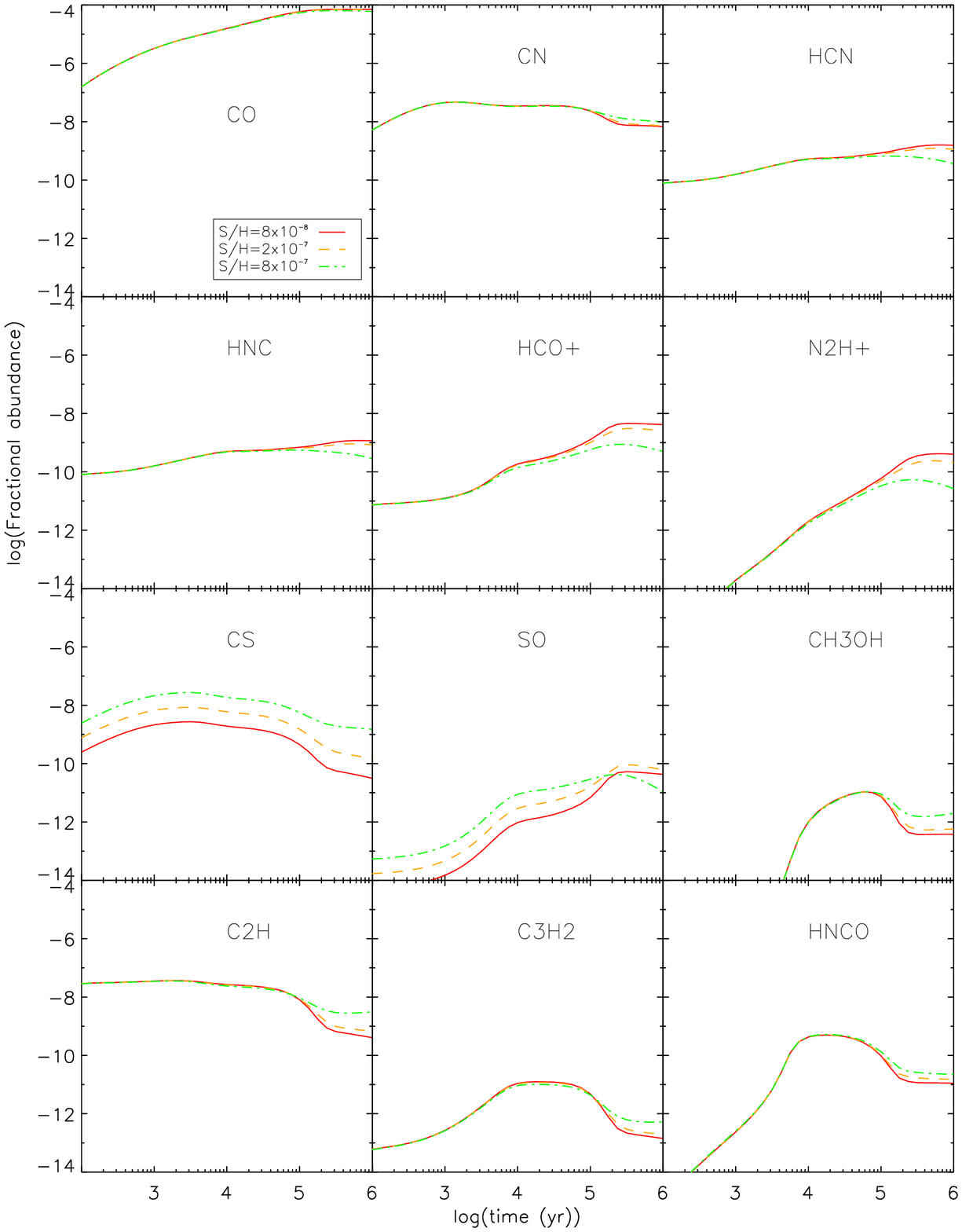}{0.9\textwidth}{}
          }
\caption{Fractional abundances of analyzed species with varying $S/H$ at $n_{\rm H2} = 3\times10^3$ cm$^{-3}$, $T_{gas} = T_{dust} = 30\,$K,
$A_{\rm V} = 2$ mag., and $\zeta = 1\times 10^{-16}$ s$^{-1}$.\label{fig:S_dep}}
\end{figure*}

The abundances of S-bearing species vary almost proportionally to the elemental abundance of sulfur.
Varying $S/H$ also affects the ionic molecules because S$^+$ can take up the charge from other ions.
 Our results show that such influence on ionized species is minor in most cases, causing  
 the decrease of molecular ions such as HCO$^+$ and N$_2$H$^+$.
 Figure \ref{fig:S_dep} shows fractional abundances in a model where the effect of $S/H$ on HCO$^+$ and N$_2$H$^+$ becomes the largest
 ($n_{\rm H2} = 3\times10^3$ cm$^{-3}$, $T_{gas} = T_{dust} = 30\,$K,
$A_{\rm V} = 2$ mag., and $\zeta = 1\times 10^{-16}$ s$^{-1}$).

\subsection{Simulated spectra}
Next, we present some of the simulated spectra from the radiative transfer calculation.
While larger fractional abundances (equivalently column densities) increase the emission intensities in general, 
the intensities also depend on the optical depths and the excitation conditions.
Molecules such as CH$_3$OH, CCH, c-C$_3$H$_2$, and HNCO stay optically thin in all of our calculation.
On the other hand, the optical depths of $^{12}$CO, CN, HCN, HNC, HCO$^+$, N$_2$H$^+$, CS, and SO
vary from an optically thin regime to $\tau \gtrsim 10$ depending on the models.
Notes on the optical depth and the effect of physical conditions on the excitation conditions are included in Appendix \ref{sec:app_exc}.

\begin{figure*}
\includegraphics[angle=270,width=.95\textwidth]{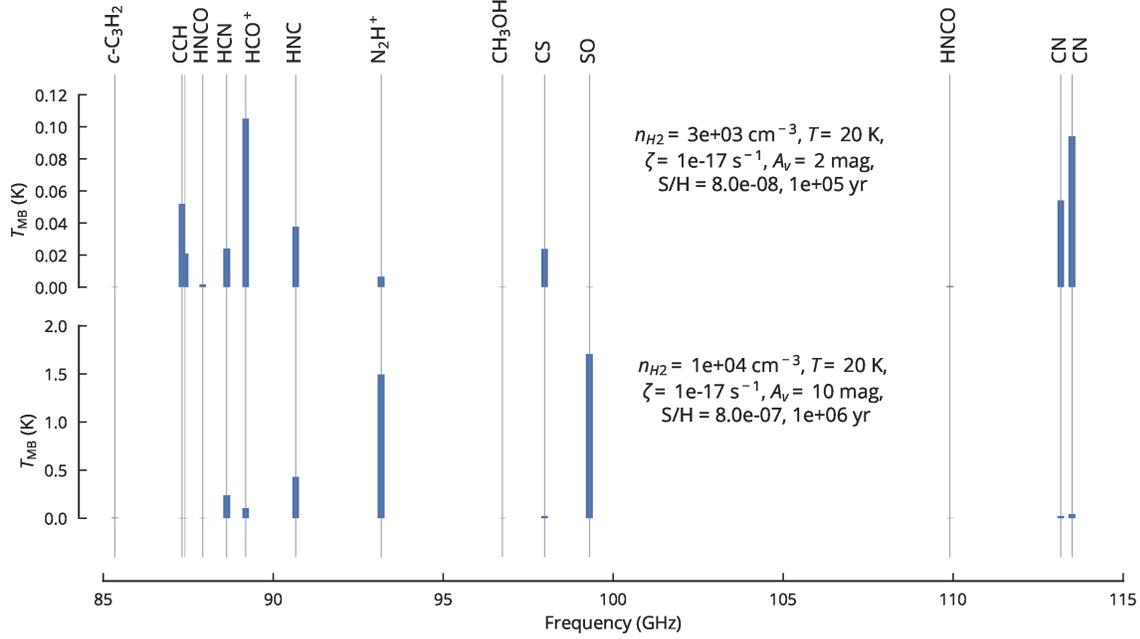}
\caption{Modeled emission peak intensities with parameters (top) $n=3\times 10^3\,$cm$^{-3}$,  $T=20\,$K,  $\zeta=1\times10^{-17}$,  $A_{\rm V}=2\,$mag, 
$S/H=8\times 10^{-8}$,  $t=1\times 10^5$\,yr as an example of a good fit with the observed spectra, and (bottom)
$n=1\times 10^4\,$cm, $T=20\,$K, $\zeta = 1\times 10^{-17}$, $A_{\rm V}=10\,$mag, $S/H=8\times 10^{-7}$, and $t=10^6\,$yr
showing the case with a large discrepancy with the observed spectra. \label{fig:spec}}
\end{figure*}

\begin{figure*}
\centering{
\includegraphics[angle=270,width=.95\textwidth]{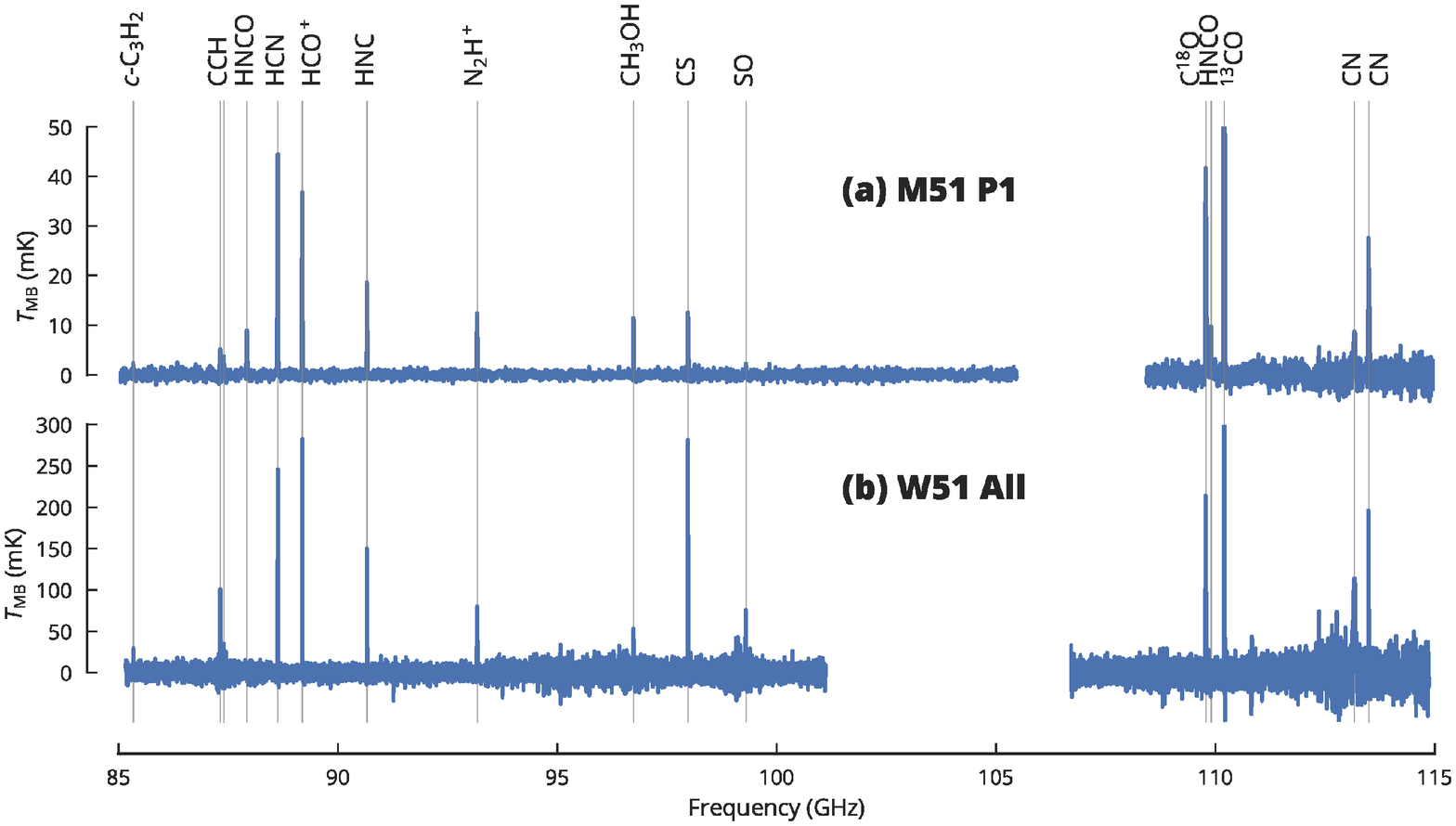}}
\caption{Observed spectra of M51 \citep{2014ApJ...788....4W} and W51 \citep{2017ApJ...845..116W} in the 3-mm band.  \label{fig:obsspec}}
\end{figure*}

In order to measure how well each model fits the observation, we calculate the correlation 
function of modeled spectra with the observed spectra of the spiral-arm region in an external galaxy M51 \citep{2014ApJ...788....4W},
 and a star-forming cloud in  our Galaxy W51 \citep{2017ApJ...845..116W}. 
The correlation coefficient was calculated as 
\begin{equation}
r= \frac{\sum \limits_{i} (I^{cal}_{i} - \overline{I^{cal}})(I^{obs}_{i} - \overline{I^{obs}})}{\sqrt{\sum \limits_{i} (I^{cal}_{i} - \overline{I^{cal}})^2 \cdot \sum \limits_{i} (I^{obs}_{i} - \overline{I^{obs}})^2}}
\end{equation}
where $I^{cal}_{i}$ is the modeled intensity and $I^{obs}_{i}$ is the observed intensity for species $i$.
We use the correlation coefficient because we aim to analyze the spectral pattern, not the absolute quantities of intensities.
For W51, we omit CN ($N=1-0, J=1/2-1/2$) from our analysis because the observed intensity seems to have a significant error\footnote{From the Einstein A coefficients, 
CN ($N=1-0, J=1/2-1/2$) is expected to have a weaker velocity-integrated intensity than CN ($N=1-0, J=3/2-1/2$). However, the observed value of CN ($N=1-0, J=1/2-1/2$) integrated intensity
is higher than that of CN ($N=1-0, J=3/2-1/2$) in W51. We investigated the reason of this trend, and concluded that the baseline subtraction of CN ($N=1-0, J=1/2-1/2$)
may not have been correct because the original data suffered from distorted baseline there.}.
Figure \ref{fig:spec} shows two examples of simulated spectra,
one with moderate fit with the observed spectra, and the other with a large discrepancy with the observe ones.
The observed spectra in W51 and M51 used for the comparison in our analysis are shown in Figure \ref{fig:obsspec}. 

\begin{figure*}
\includegraphics[width=0.9\textwidth]{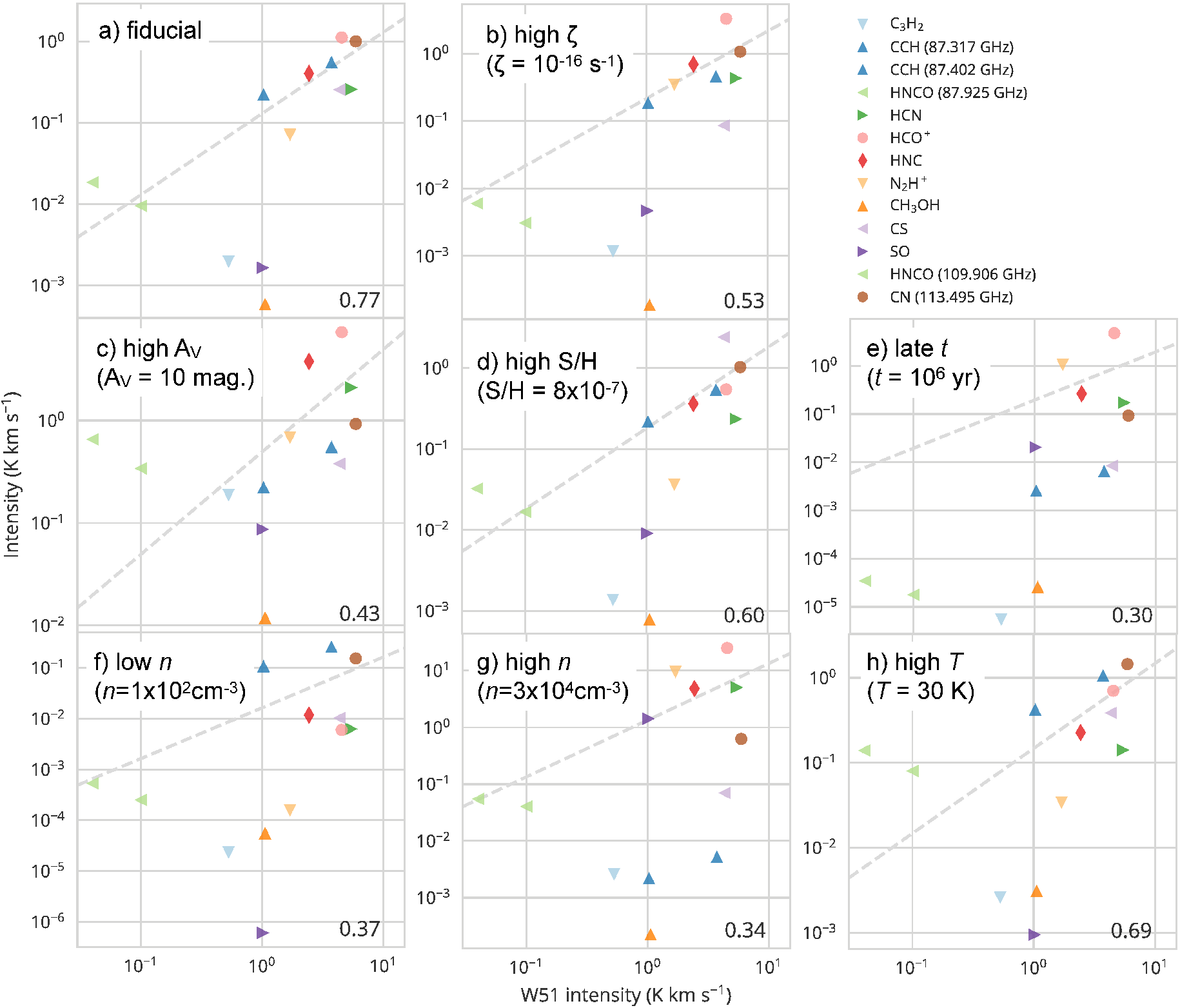}
\caption{The abscissa shows observed intensities of W51, and the ordinate shows the modeled intensities
with parameters shown above each panel.  
The fiducial model shown in Panel a) has physical parameters of $n=3\times 10^3\,$cm$^{-3}$, $T=20\,$K, $\zeta = 1\times 10^{-17}\,$s$^{-1}$,
$A_{\rm V}=2\,$ mag., $S/H=8\times 10^8\,$, and $t=10^5\,$yr.
For other panels, only the physical parameters that are different from the ones of the fiducial model are shown.
A correlation coefficient for each model is shown at right bottom of each panel. \label{fig:spec_W51}}
\end{figure*}

\begin{figure*}
\includegraphics[width=0.9\textwidth]{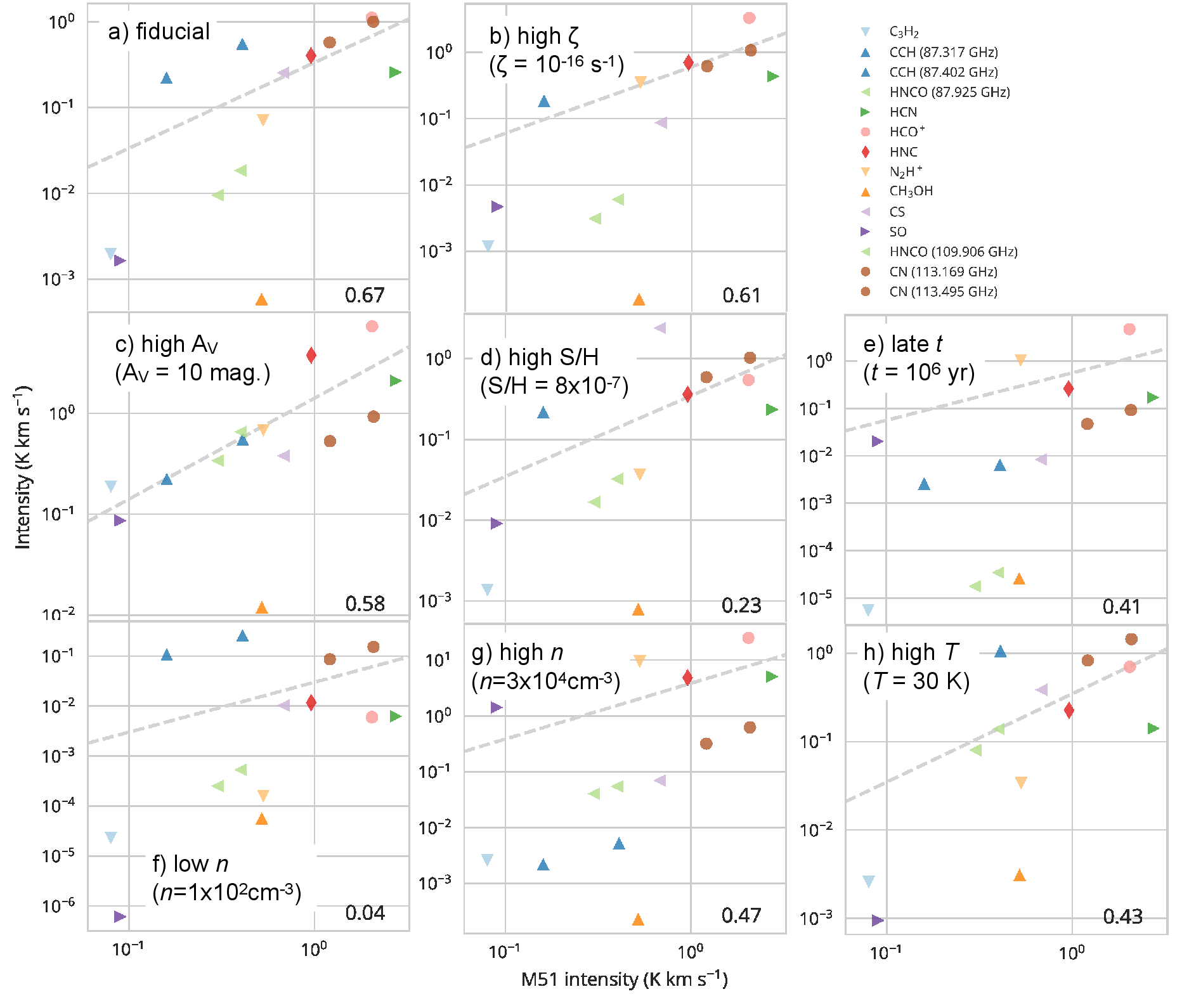}
\caption{The same as Figure \ref{fig:spec_W51}, but the abscissa shows the observed intensities of M51. \label{fig:spec_M51}}
\end{figure*}

Quantitative comparison between models and observations for each species is made in Figures \ref{fig:spec_W51} (comparison with W51)
and \ref{fig:spec_M51} (comparison with M51). 
Here we choose the fiducial model from one of the highest $r$ with the parameters $n=3\times 10^3\,$cm$^{-3}$,  $T=20\,$K,  $\zeta=1\times10^{-17}$,  $A_{\rm V}=2\,$mag., $S/H=8\times 10^{-8}$,  $t=1\times 10^5$\,yr, and discuss the variation from the fiducial case 
when the parameters are varied (Panel a in Figures \ref{fig:spec_M51} and \ref{fig:spec_W51}).
The major change with the increased $\zeta = 10^{-16}\,$s$^{-1}$ is the increased HCO$^+$ and N$_2$H$^+$ intensities due to the abundance change (Panel b). 
If the visual extinction is increased from $A_{\rm V} =2$\,mag to 10 mag, CCH and CN intensities decrease,
but this change is favorable to the fit to the observations (Panel c). Meanwhile, HCO$^+$, HCN and HNC intensities increase with
increased $A_{\rm V}$.
Increased $S/H$ causes the CS emission to be overproduced compared with observed values because the abundances of sulfur species increase (Panel d).
Sulfur monoxide is not overproduced in low-density cases, but it is overproduced in high-density cases.
If we take the intensities at later time ($t=10^6\,$yr) than the fiducial case, then intensities of HCO$^+$ and 
N$_2$H$^+$ become higher. Changes in both intensities can be accounted for from higher abundances (Panel e).
For the high-density cases, the emission intensities of HCO$^+$ become overproduced in the model.
When the density is lower, the CCH and CN intensities increase (Panel f). On the other hand, higher densities enhance the intensities of 
HCO$^+$, N$_2$H$^+$, and SO (Panel g).
Higher temperatures do not affect most species, except for CH$_3$OH and HNCO (Panel h).
The intensities of CH$_3$OH is underproduced in all the cases. This underproduction of gas-phase CH$_3$OH may result from the turbulence 
or shock that causes the increase of methanol abundances in observations of molecular clouds.
If the intensities of CH$_3$OH are enhanced through a mechanism that we did not include,
our results would not be reasonable if this deviation of CH$_3$OH from the observations affects the correlation coefficients significantly. 
Therefore, we also tried calculation of correlation coefficients without CH$_3$OH,
and confirmed that correlation coefficients with and without CH$_3$OH only change at most 0.05, on average, by 0.01-0.02.
Furthermore, the change caused by excluding CH$_3$OH affect most models in a similar way, 
and the discussion of constraints of physical conditions in following sections is essentially unchanged whether we calculate 
correlation coefficients with or without CH$_3$OH. Our analysis include methanol in the derivation of correlation coefficients.

\section{Discussion}\label{sec:disc}
\subsection{Constraints on physical conditions}\label{sec:bestfit}

\subsubsection{Constraints from W51}\label{sec:const_w51}
\begin{figure*}
\includegraphics[width=.99\textwidth]{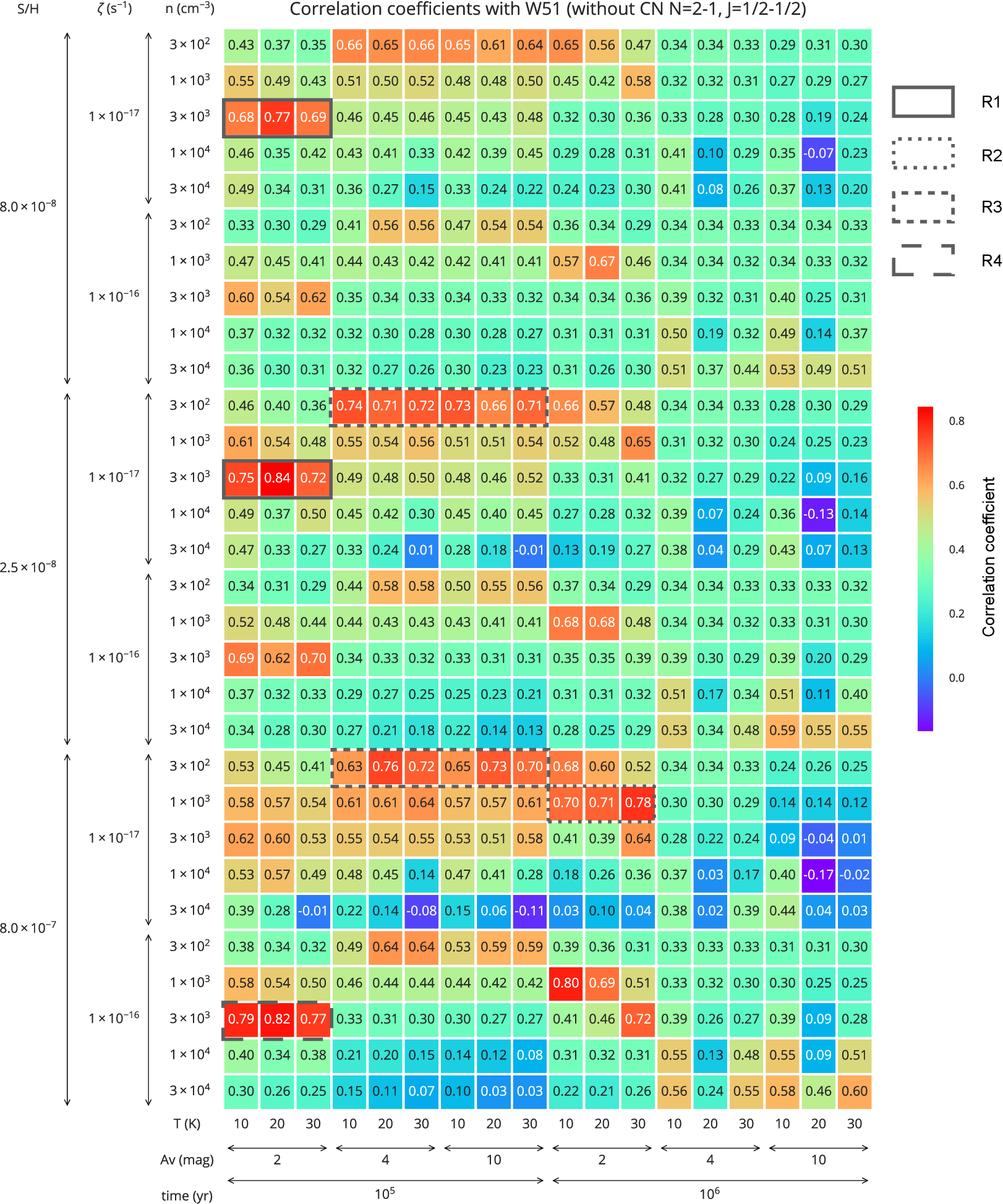}
\caption{Correlation coefficients between modeled emission intensities and observed intensities of W51.
Sets of parameters with moderate agreement are shown with grey rectangles. \label{fig:W51coeff}}
\end{figure*}

Figure \ref{fig:W51coeff} shows correlation coefficients for sets of the physical parameters for W51.
There are 4 regions in the parameter space that have relatively high correlation coefficients ($r \gtrsim 0.7$),
and those regions are indicated as grey rectangles in Figure \ref{fig:W51coeff}.
The first region is $n=3\times 10^3\,$cm$^{-3}$, $A_{\rm V}=2\,$mag with $S/H = (0.8-2.5)\times 10^{-7}$, $t=10^5\,$yr and $\zeta = 10^{-17}\,$s$^{-1}$
(Parameter Space Region (R) 1). 
The second region is $n=1\times 10^3\,$cm$^{-3}$, $A_{\rm V}=2\,$mag, $\zeta = 10^{-17}\,$s$^{-1}$ and $t=10^6\,$yr (R2).
The third region indicated as R3 has parameters of $n=3\times 10^2\,$cm$^{-3}$, $A_{\rm V}=4-10\,$mag, $\zeta = 10^{-17}\,$s$^{-1}$ and $t=10^5\,$yr, 
and $S/H = (2.5-8)\times 10^{-7}$, and the last region, R4, has physical parameters of 
$n=3\times 10^3\,$cm$^{-3}$, $A_{\rm V}=2\,$mag with $S/H = 8\times 10^{-7}$, $t=10^5\,$yr and $\zeta = 10^{-16}\,$s$^{-1}$.
We choose those regions because they have high correlation coefficients for relatively large volume in the parameter space.
Although some sets of physical parameters show high correlation coefficients, we do not consider them to be likely conditions
if the correlation coefficients are low in almost all of their neighboring locations in the parameter space.

\subsubsection{Constraints from M51}
\begin{figure*}
\includegraphics[width=.99\textwidth]{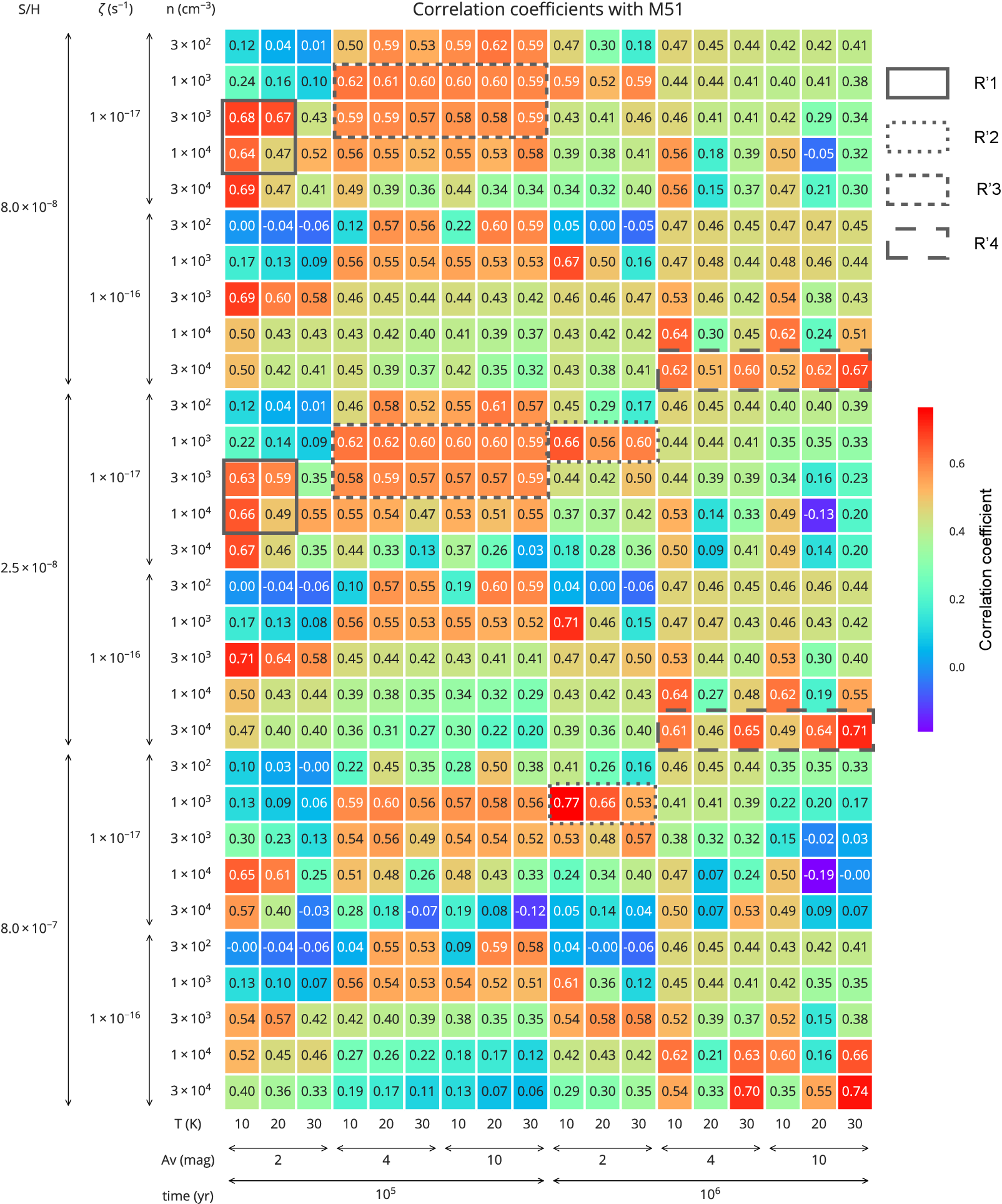}
\caption{Same as Figure \ref{fig:W51coeff}, but for M51. \label{fig:M51coeff}}
\end{figure*}

For the case of M51, regions similar to R1, R2, and R3 in W51 also have relatively good agreement, and they are marked as R$'$1, R$'$2, and R$'$3.
However, compared with R1 and R3, R$'$1 and R$'$3 tend to have either higher densities.
For example, R$'$1 in Figure \ref{fig:M51coeff} has a higher density range of $n=(3-10)\times 10^3\,$cm$^{-3}$ compared with R1 ($n=(1-3)\times 10^3\,$cm$^{-3}$). 
The temperature range is also slightly narrower in R$'$1 than in R1 ($T=10-30\,$K in R1 while $T=10-20\,$K in R$'$1). 
Likewise, R$'$3 has agreement with $n=(1-3)\times 10^3\,$cm$^{-3}$ instead  $n=3\times 10^2\,$cm$^{-3}$ in R3.
Another difference between R3 and R$'$3 is the range of sulfur elemental abundance, which is lower in R$'$3 ($S/H = (0.8-2.5)\times 10^{-7}$).

In addition to regions R$'$1-3, which are similar to R1-3, there is a parameter space region R$'$4 where the agreement is relatively high for the case of M51 (Figure \ref{fig:M51coeff}).
Physical parameters of R$'$4 are $n= 3\times 10^4\,$cm$^{-3}$, $A_{\rm V} = 4-10$, and $t=10^6\,$yr.
However, we argue that they are unlikely to represent the real physical conditions.
Despite some differences, spectra from M51 and W51 are similar, 
and the correlation coefficient between spectra of M51 and W51 shows a relatively high value of 0.83.
Therefore, the true physical parameters in those observations should be close to each other.
This means that physical parameters that show high correlation coefficients for both sources are 
more likely solutions. If this is the case, R$'$4 can be excluded from the likely physical parameters.
There is another reason to exclude R$'$4 from the plausible parameter space; the high correlation with the observations in R$'$4
is likely to be caused by a non-physical behavior of the chemical model.
As discussed in \citet{2007A&A...467.1103G}, at late times, CO and CH$_3$OH can be converted to CH$_4$
due to cosmic-ray induced photodissociation of CH$_3$OH into CH$_3$. Methane can be efficiently formed from CH$_3$ on the grain surface.
With time, methane can be converted into other hydrocarbons. However, such behavior in the late time has not been well understood,
and has not been observed in the astrophysical environment.
In our modeled results, CH$_4$ is indeed overproduced at 10 K within the physical conditions of R$'$4,
and becomes more abundant than CO ice. Observationally, CH$_4$ is seen to be less abundant than CO ice \citep[][]{2011ApJ...740..109O}. 
On the other hand, for the case of 30 K, CH$_4$ is not the most abundant form of ice; CH$_4$ is volatile, and the majority of CH$_4$ is in the gas-phase.
Instead of CH$_4$, CO$_2$ and C$_2$H$_6$ takes over as abundant forms of ice. 
Abundant C$_2$H$_6$ still leads to a similar effect as the abundant CH$_4$.

\subsubsection{Constraints from absolute intensities}
In above discussion, we compare the spectral pattern of models and observations,
 but not the brightness temperature itself. This is because high-density clouds
are likely to have low volume filling factors in molecular-cloud averaged spectra. Since the gas with higher densities gives higher intensity 
for a given column density, it is possible for models to be consistent with observations even if modeled intensities are higher than observed values.
However, we can exclude some models if they underproduce intensities 
 even when maximum possible total column densities are used for radiative transfer calculations.
 Because of this reason, we argue that molecular emission from $n=3\times10^2\,$cm$^{-3}$ is not the dominant contributor to our observed spectra.
 Since our observation of W51 covers a 50-pc region, the maximum possible total column density from gas with $n=3\times10^2\,$cm$^{-3}$ 
 is $N_{\rm H2} \sim 5\times10^{22}\,$cm$^{-3}$. When we run models with this total column density, emission predicted from most models fails to produce 
 high enough intensities equivalent to the observed values for the case of $n=3\times10^2\,$cm$^{-3}$. 
 A few lines in limited models do produce as much intensities as the observed ones,
 but those models do not show low correlation coefficients when we compare their spectral pattern with the observed one.

\subsubsection{Constraints from CI/CO intensity ratios}\label{sec:const_cico}
\begin{figure*}
\includegraphics[width=.99\textwidth]{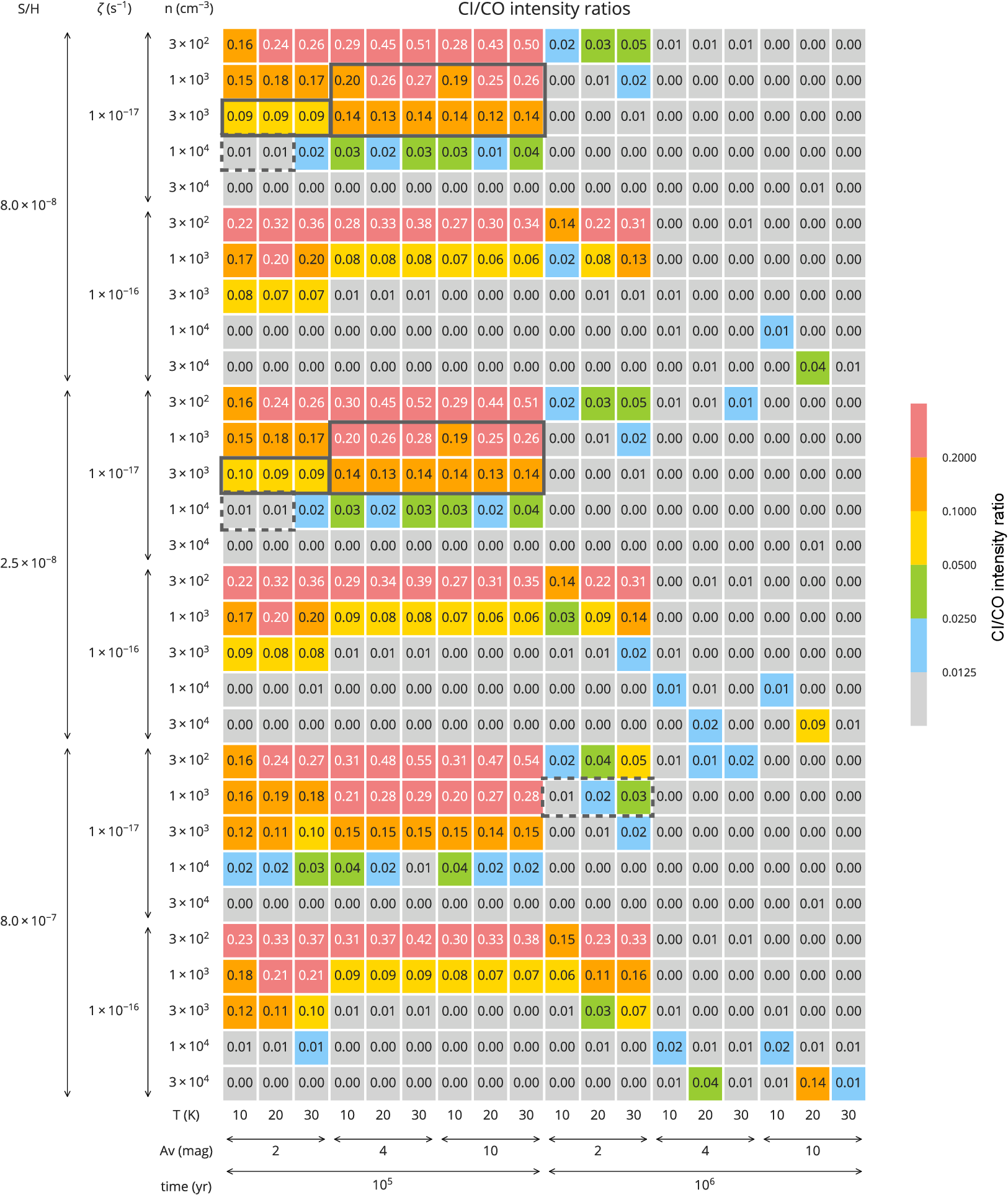}
\caption{Intensity ratios of CI/$^{12}$CO(1-0) in the grid of physical parameters.
Constraints obtained from W51 and M51 observations, and from absolute intensities are shown as solid or dotted grey rectangules \label{fig:CI-CO}}
\end{figure*}

From constraints obtained from W51, M51, and the absolute intensity argument above, we have sets of physical parameters with moderate agreement with observations.
In order to impose another constraint, we also simulated the intensities of CO($J=1-0$) and CI($^3P_1$ - $^3P_0$).
The simulated CI/CO intensity ratios are shown in Figure \ref{fig:CI-CO}.
Although CI/CO ratios toward our target areas of W51 and M51 are not reported, CI/CO ratios in molecular clouds are generally 
known to be 0.05-0.1 \citep{2002ApJS..139..467I,2003ApJ...589..378K}.
This CI/CO ratio is observationally found to be relatively constant in various regions of molecular clouds.
Thus, we compare this ratio with our model.
Parameters that have moderate agreement with observations from constraints obtained above are also shown as grey rectangles with solid lines or dotted lines.
If we add another constraint of a factor of 4 within this range $0.0125< {\rm CI/CO} < 0.4$,
most cases at $t=10^6\,$yr are excluded. 

From all the constraints from Sections \ref{sec:const_w51}-\ref{sec:const_cico}, we conclude that sets of physical parameters that are likely representing conditions in observations 
are regions shown as solid grey rectangles in Figure \ref{fig:CI-CO}.
Those sets of parameters are $n=3 \times 10^3\,$cm$^{-3}$, $A_{\rm V}=2\,$mag, $\zeta = 10^{-17}\,$s$^{-1}$, $S/H = (0.8-2.5)\times 10^{-8}$, and $t=10^5\,$yr,
or $n=(1-3) \times 10^3\,$cm$^{-3}$, $A_{\rm V}=4-10\,$mag, $\zeta = 10^{-17}\,$s$^{-1}$, $S/H = (0.8-2.5)\times 10^{-8}$, and $t=10^5\,$yr.

\subsection{Density constraints}
The range of densities constrained by our model is $n=(1-3)\times 10^3\,$cm$^{-3}$, which is much lower than 
the critical densities of most of the species in the optically thin case \citep[see Table 3 of ][ for values of critical densities]{2017ApJ...848...17N}. 
This range of densities is still reasonable in the optically thick case.  
The critical densities considering the optical depth by \citet{2005pcim.book.....T} is lower by a factor of 50 
at $\tau = 10$ compared with the optically thin case due to radiation trapping.
 The optical depth in our models vary as discussed in Appendix \ref{sec:app_exc},
and there are cases where the optical depth is as high as 10 or more.
\citet{2015PASP..127..299S} calculated the effective  critical density in the optically thick cases.
If the column densities of the species are $N=10^{14}\,$cm$^{-2}$, similar values to those in our models,
HCO$^+$ and HCN have effective critical densities of $10^3$ and $8\times 10^3\,$cm$^{-3}$, respectively.

\citet{2017A&A...605L...5K} derived the mean density of gas in their Orion A observations.
In this analysis, they assumed a power-law density distribution of the filament, and made a fit to the dust
column density. As a result, they concluded that mean density of line-emitting gas is 870\,cm$^{-3}$ for
most molecules ($^{12}$CO, $^{13}$CO, C$^{18}$O, CN, CCH, and HCN),
while N$_2$H$^+$ is emitting from gas with the density $n \sim 4000\,$cm$^{-3}$.
Although we used the averaged spectra in the larger observed area for our analysis, 
our results show a similar range of density to the one suggested by \citet{2017A&A...605L...5K}.
\citet{2017A&A...599A..98P} also derived similar values of densities in their Orion B observations;
  in their regions sub-divided by $A_{\rm V}$, lines of most species are emitted from regions with mean densities of $n=890\,$cm$^{-3}$ and $n=2700\,$cm$^{-3}$ 
except for CH$_3$OH, H$^{13}$CO$^+$, and N$_2$H$^+$, whose emission also comes from 
higher density regions of $n=7500\,$cm$^{-3}$. 
Both \citet{2017A&A...605L...5K} and \citet{2017A&A...599A..98P} observed smaller area than \citet{2017ApJ...845..116W} did in W51,
yet, their derived densities are equivalent to our best-fit densities.
Our results provide another evidence that the low-density gas can generate high enough luminosity for the species
that are conventionally thought as ``dense-gas tracers."

\subsection{Remarks on visual extinction}\label{sec:vis-ex}
In our model analysis (Section \ref{sec:bestfit}), there are two different ranges of $A_{\rm V}$: $A_{\rm V}=2\,$mag in R1 and $A_{\rm}=4-10\,$mag in R2. 
In our observed area, there are variations of densities, temperatures, and visual extinctions.
For the case of the density, components with relatively low density is the main contributor to the average molecular emission as discussed above.
However, a relatively good fit over various values of $A_{\rm V}$ may suggest that the contribution from various values of $A_{\rm V}$ may be comparable. 
Therefore, multi-component models with different values of $A_{\rm V}$ may give a better fit to observations.

Having good agreement over a wide range of visual extinction also seems to be reasonable when we consider other observational data.
\citet{2017A&A...599A..98P} reported that 3, 51, 40, and 8 \% of the CO-traced gas mass is distributed in regions with 
$1<A_{\rm V} ({\rm mag})<2$, $2<A_{\rm V} ({\rm mag})<6$, $6<A_{\rm V} ({\rm mag})<15$, and $15<A_{\rm V} ({\rm mag})<222$, respectively,
while the ratios are 8, 38, 36, and 20 \% when the mass is traced by dust. The interstellar radiation field $G_{0}$ in their observations has 
a wide range of values of 4 - 28,000, with the mean values of individual regions ranging from 30 to 72.
Note that we used $G_{0}=1$ in our model, and $A_{\rm V}=2\,$mag in our model corresponds to the chemistry
with $A_{\rm V}=3.1 - 4.1\,$mag for the above-mentioned mean radiation field of $G_{0}= 30-72$ (assuming $\gamma = 2-3$; see footnote \ref{ftn:photodis} in Section \ref{sec:fracabun}).
Similarly, the chemistry in $A_{\rm V}=4\,$mag in $G_0=1$ corresponds to $A_{\rm V}=5.1 - 6.1\,$mag when $G_{0}= 30-72$.
\citet{2017A&A...605L...5K} reported $A_{\rm V}$ of 6.1\,mag for most species and $A_{\rm V}$ of 16\,mag for N$_2$H$^+$
although they did not list the range of $G_0$.

\subsection{Implication of the timescale}
The simulated spectra and the CI/CO ratio suggest that the chemical time of $10^5\,$yr is preferred in our model.
This time scale of $t=10^5\,$yr is much shorter than the lifetime of molecular clouds \citep[$10^6 - 10^7\,$yr; e.g.,][]{2001ApJ...562..852H,2004ApJ...616..283T}.
The free fall time of clouds at the densities $n=(1-3)\times 10^3\,$cm$^{-3}$ is $(0.6 - 1)\times 10^6\,$yr,
which is still longer than our chemistry time. 
On the other hand, the turbulent crossing time $t_{\rm cross}$ is $\sim 5\times 10^5 (\frac{R}{0.1\,{\rm pc}}) (\frac{v_{\rm turb}}{0.2\, {\rm km~s}^{-1}})^{-1}$, 
where $R$ is the length and $v_{\rm turb}$ is the turbulent velocity.
The total column density that we used in the model ($N=10^{22}\,$cm$^{-2}$) corresponds to the source size of $1-3\,$pc for $n = (1-3)\times 10^3\,$cm$^{-3}$,
but the actual geometry of the source is likely to be filamentary with the width of $R \sim 0.1\,$pc.
For $R=1\,$pc, $v_{\rm turb}$ is 1.1 km s$^{-1}$ by assuming the object follows Larson's law \citep{1981MNRAS.194..809L}, which gives $t_{\rm cross} $ of $10^6$\,yr.
If $R=0.1\,$pc, the Larson's Law gives $v_{\rm turb} $ of $0.45\,$km s$^{-1}$, which means the crossing is $2.2 \times 10^5\,$yr.
This is equivalent to, in order of magnitude, our derived chemistry time.
If such turbulence occurs, molecular medium may be constantly exposed to low-density environment with high flux of UV photons to dissociate molecules.
 If $G_0 = 30$, $A_{\rm V} \ll 1$ and the rate coefficient is a typical value of $k = 10^{-10}$, photodissociation can easily occur in a very short timescale of 10 yrs. 
This photodissociation can bring the chemistry close to the initial condition of atomic/ionic state, and setting the chemistry clock back to zero.
It should be noted that our models were run with the pseudo-time-dependent approach, 
and the actual age of the molecular cloud may by longer than our derived chemistry time scale.

\subsection{Excitation with electrons}
\sloppypar{In regions with high electron fractions, collisional excitation with electrons can be important 
in addition to the ones with H$_2$ or H.
This electron excitation was in fact brought up as a possible mechanism of enhancing HCN emission intensities
in strongly irradiated regions. 
\citet{2017ApJ...841...25G} analyzed the effect of electron excitation, and concluded that this excitation mechanism may be significant when $X$(e$^-$)$\gtrsim 10^{-5}$
and $n \lesssim 10^{5.5}\,$cm$^{-3}$ for the case of HCN.}
In our chemical model, there are some conditions where the electron excitation becomes important.
The highest electron fraction is $X$(e$^-$) $\sim 2.9\times 10^{-5}$ when $n=3\times 10^2\,$cm$^{-1}$, $A_{\rm V}=2\,$mag and $\zeta=10^{-16}\,$s$^{-1}$.
When $n=1\times 10^3\,$cm$^{-3}$, there are still cases where the electron fraction is high enough for collisions with electrons to be significant ($X$(e$^-$) $\sim 1.6\times 10^{-5}$).
For the higher density of $n = 3 \times 10^3\,$cm$^{-3}$, $X$(e$^-$)  $\lesssim 2\times 10^{-6}$ even with a set of parameter to cause the highest ionization fraction ($A_{\rm V}=2\,$mag, $\zeta=10^{-16}\,$s$^{-1}$,
and $S/H = 8\times 10^{-7}$), and the electron excitation gives only a minor contribution to the emission.
 Unfortunately, the collisional cross section with electron is not available for all the species, which are needed for the comparison of the spectrum. 
For this reason, we do not consider the electron excitation here, but the results of low-density cases need to be taken with caution.

\subsection{Two-component model}
\begin{figure*}
\includegraphics[width=.99\textwidth]{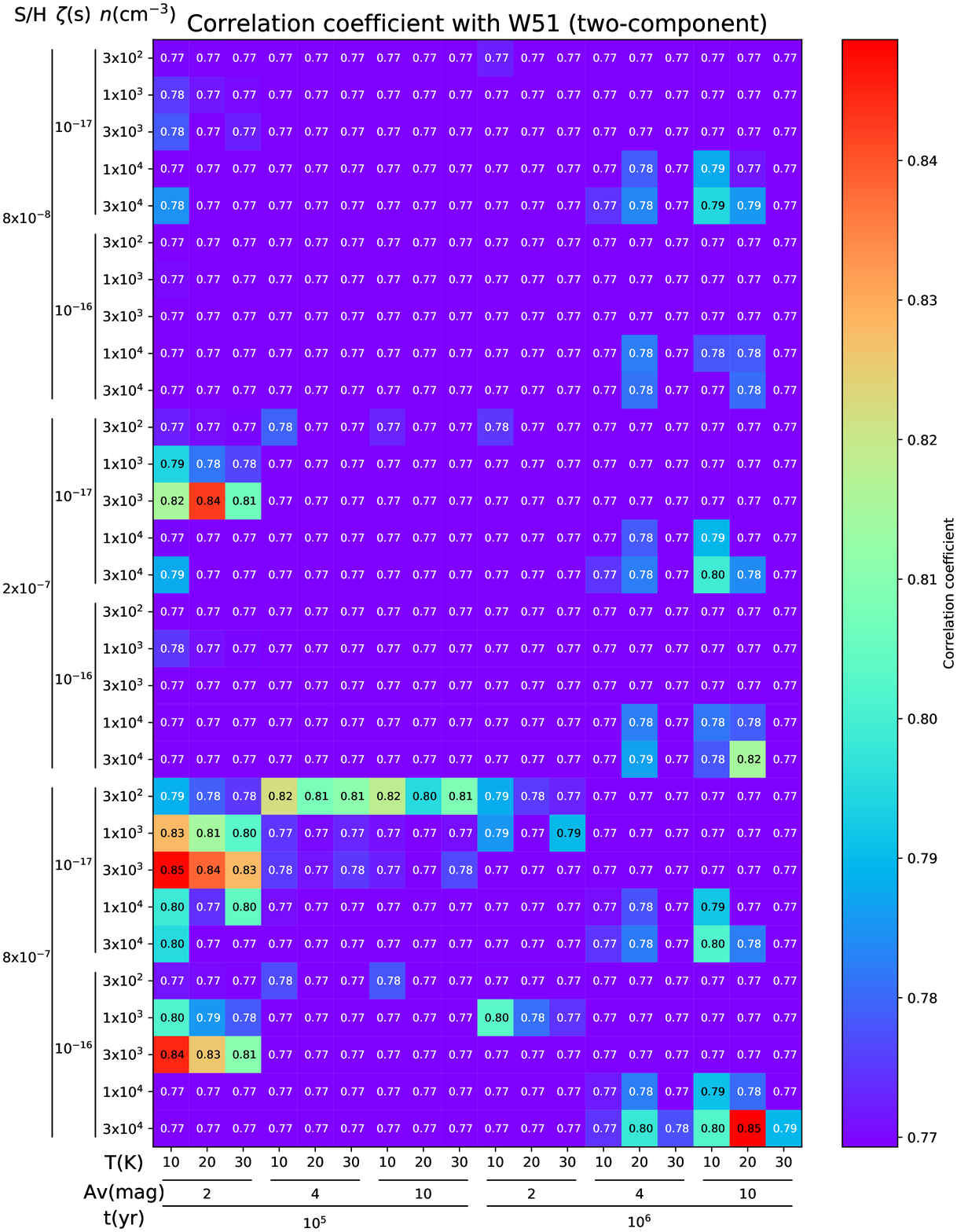}
\caption{Correlation coefficients between modeled emission in two-component model and observed intensities in W51.
For the two-component model, the emission intensities of each model is added to the fiducial model with varying fractions from 0-100\%,
and the maximum correlation coefficient among various fractions is shown for each model. \label{fig:2comp-w51}}
\end{figure*}

\begin{figure*}
\includegraphics[width=.99\textwidth]{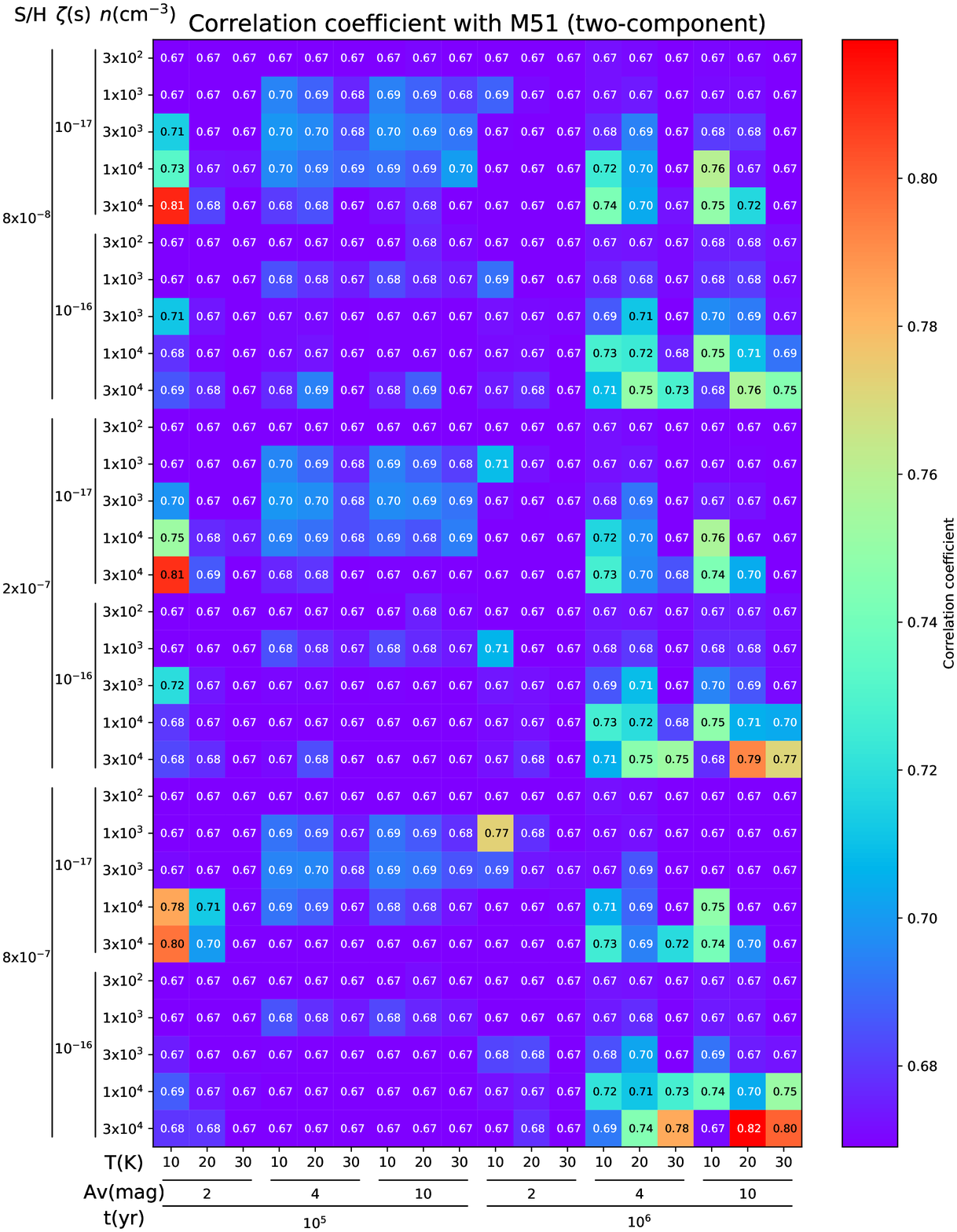}
\caption{Same as Figure \ref{fig:2comp-w51}, but for M51. \label{fig:2comp-m51}}
\end{figure*}

In Section \ref{sec:vis-ex}, we argue that the molecular emission likely comes from multiple components. 
To test this claim, we examine whether the superposition of two components provides better fit than the single-component model.
We found that better agreement with observations is achieved for M51, but not significantly for W51.
To create two-component spectra, we added modeled spectra of each model to that of the fiducial model with varying fraction of 0-100\% by 1\% increments.
Then, we calculate the correlation coefficients between the observations and modeled spectra.
The maximum value of correlation coefficients among models of varying contributions from the fiducial model is kept for each model.
Those values are shown in Figures \ref{fig:2comp-w51} (W51) and \ref{fig:2comp-m51} (M51).
In those figures, only the physical parameters of the additional components are shown.
From the way we construct the two-component models, the minimum values of correlation coefficients in the two-component model are ones of the fiducial model.
The improvement of using two-component models is not significant for the case of W51. The highest improvement is made by combining with the model with higher $S/H$ to the fiducial model,
not necessarily the models with different physical parameters.
There are still several cases where the high-density ($n=3\times 10^4\,$cm$^{-3}$) and late-time ($t=10^6\,$yr) model gives better correlation coefficients than fiducial model alone.
On the other hand, for the case of M51, correlation coefficients improve by more than 0.1 when additional components of high-density ($n=(1-3)\times 10^4\,$cm$^{-3}$) models are combined with the fiducial model. 
Among them, high correlation coefficients are achieved at a low $A_{\rm V}$ of 2 mag and $t=10^5\,$yr, or at higher $A_{\rm V} = 4-10\,$mag and $t=10^6\,$yr.

The use of two-component model also improves the disagreement of SO/CS ratio in some models. 
Observed ratios of SO/CS are 0.13 in M51 and 0.23 in W51. However, this ratio is underproduced by a few orders of magnitude in most models that have high correlation coefficients 
in single-component models, as shown in Figures \ref{fig:spec_W51} and \ref{fig:spec_M51}.
In two-component models, there are a few models that reproduces the observed SO/CS ratios with in a factor of a few while having high overall correlation coefficients.

\section{Summary}
In this paper, we model the emission intensities of molecular species commonly observed in the 3-mm band
using the grid of physical parameters. We conducted molecular abundance calculations using the gas-grain time-dependent chemical model followed by
radiative transfer calculations.
Our results are compared with observations taken at a few tens of parsec scale in W51 and M51.

Below we list our main findings.
\begin{itemize}
\item Variation of emission intensities with physical parameters can be mostly explained by abundance variations.
The dependence of modeled intensities on physical parameters and comparison with observed spectral pattern are summarized as follows. 
\begin{itemize}
\item When $A_{\rm V} =2\,$mag in higher densities, N$_2$H$^+$ is formed in shorter time, and its intensity increases.
\item The emission intensities of HCO$^+$ and N$_2$H$^+$ increase with the higher cosmic-ray ionization rate. 
\item N$_2$H$^+$ and HCO$^+$ intensities are higher at $t=10^6\,$yr than intensities at $t=10^5\,$yr.
\item With higher visual extinction, emission intensities of CN and CCH become weaker while those of HCO$^+$, N$_2$H$^+$ and HNC increase.
\item The emission intensities of sulfur-bearing molecules CS and SO increase with the higher elemental abundance of sulfur.
The higher sulfur abundance overproduce CS intensities,
while SO is overproduced in some high-density, late-time model when compared with observed spectral pattern.
\item Our models do not reproduce enough intensities of CH$_3$OH in all the cases. This underproduction suggests that CH$_3$OH 
may be enhanced by a mechanism that are not included in our model such as sputtering through shocks or turbulence.
\end{itemize}

\item Our comparison of models with the observations of W51 and M51 suggests the physical conditions of $n=(1-3)\times 10^3\,$cm$^{-3}$, 
$A_{\rm V} \geq 4\,$mag, $\zeta = 10^{-17}\,$s$^{-1}$, $t=10^5\,$yr
or $n=3 \times 10^3\,$cm$^{-3}$, $A_{\rm V} =2\,$mag, $\zeta = 10^{-17}\,$s$^{-1}$, $t=10^5\,$yr, and $S/H=(0.8-2.5)\times 10^{-7}$.
\item The derived values of the density are lower than critical densities of most species in the optically thin case, but are similar to observationally determined values in Orion A and Orion B clouds \citep{2017A&A...605L...5K,2017A&A...599A..98P}. Our results provide another supporting evidence that enough emission can come from relatively low-density regions.
\item The short timescale derived in our analysis ($t=10^5\,$yr) is consistent with the turbulent mixing timescale that allows exposing the medium to UV-photons, which refreshes the chemistry clock.

\item Better correlation with observations of M51 is achieved by two-component models, i.e., summation of spectral pattern of some of high-density models to that of the fiducial model for the case of M51.  
For the case of W51, improvement of the correlation is not significant. Such use of two-component models can also alleviate the discrepancy between observation and models for sulfur-bearing
molecules, CS and SO.
\end{itemize}

The observed spectral pattern and the derived physical condition from the model serves as a benchmark of the chemistry in a relatively quiescent molecular cloud.
In future studies of regions with more extreme physical conditions, the differences from this benchmark can be examined.

\acknowledgments

We thank the anonymous referee for the thorough reading of the manuscript and thoughtful comments.
We are grateful to Ugo Hincelin for sharing his network with us. NH is supported by a grant from Ministry of Science and Technology in Taiwan MOST 107-2119-M-001-041- and MOST 107-2119-M-001-022-.
SY, YA, and NS acknowledge the financial support by JSPS KAKENHI Grant number 18H05222.
YN was supported by NAOJ ALMA Scientific Research grant Number 2017-06B 
and JSPS KAKENHI Grant Number JP18K13577.

\software{RADEX \citep{2007A&A...468..627V}, Nautilus \citep{2009A&A...493L..49H,2010A&A...522A..42S}}

%






\appendix

\section{Notes on the excitation conditions}\label{sec:app_exc}
\begin{figure*}
\vspace{-12mm}
\gridline{\fig{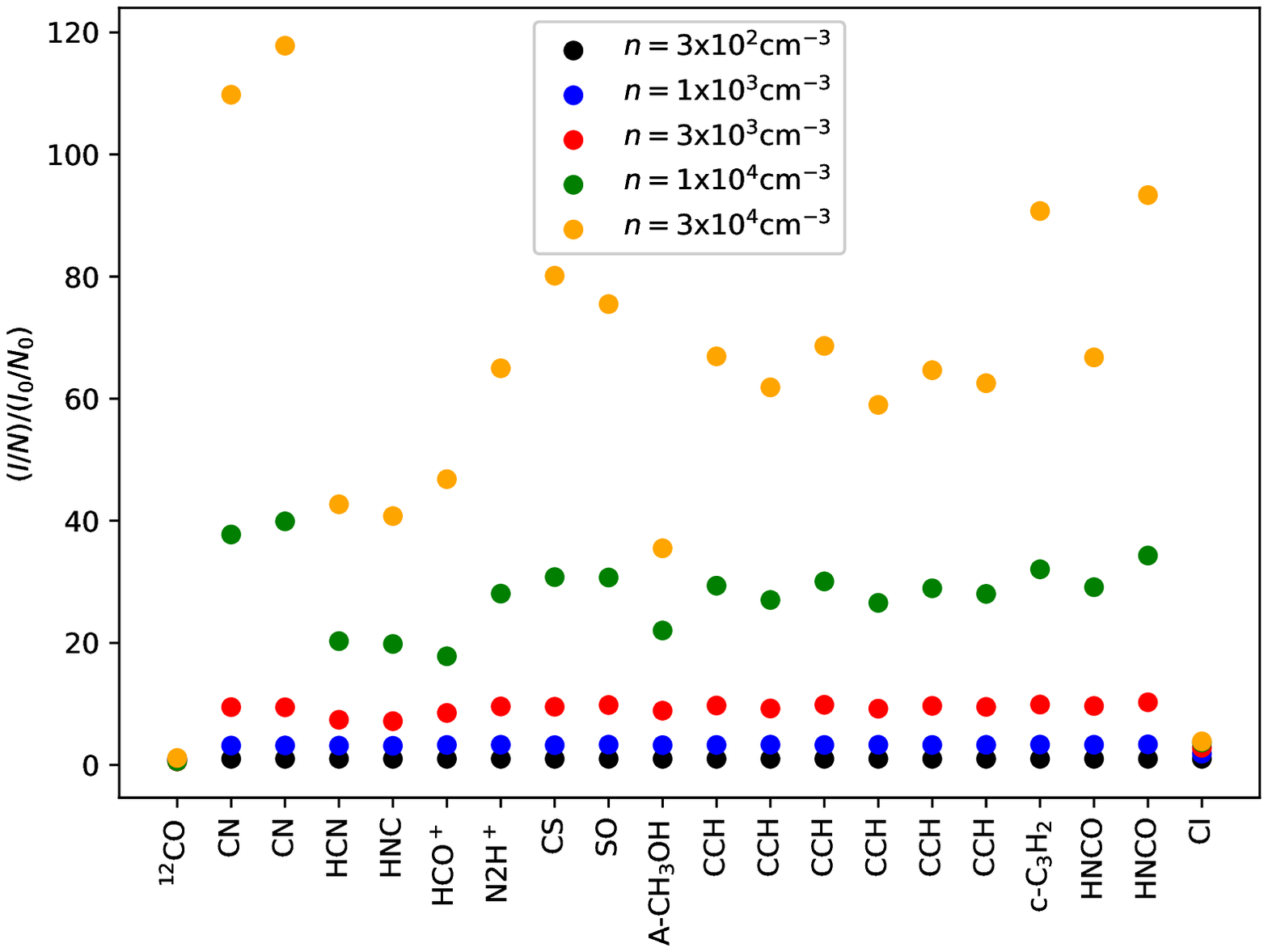}{0.45\textwidth}{ \vspace{40mm}(a)}
\fig{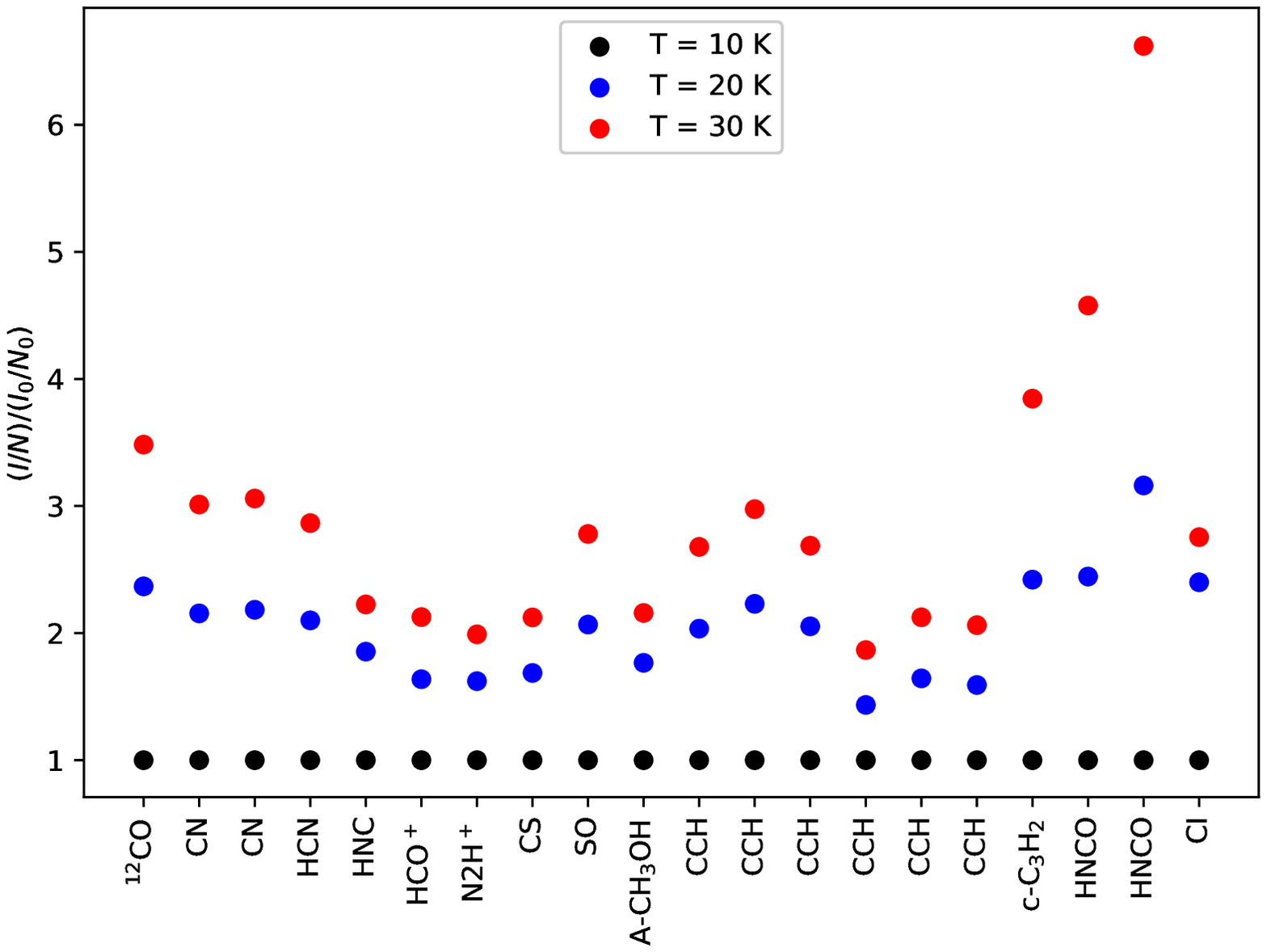}{0.45\textwidth}{ \vspace{40mm}(b)}
}
\vspace{2mm}
\caption{Modeled intensities divided by column densities
($left$) with varied densities normalized to the case of $n=1\times 10^3\,$cm$^{-3}$ with $A_{\rm}=2$\,mag, $\zeta = 10^{-17}\,$s$^{-1}$, $S/H=8\times 10^{-8}$, and $T=10\,$K
($right$) with varied temperatures normalized to the case of $T=10\,$K with $A_{\rm}=2$\,mag, $\zeta = 10^{-17}\,$s$^{-1}$, $S/H=8\times 10^{-8}$, and $n=3\times 10^{3}\,$cm$^{-3}$. \label{fig:intcol}}
\end{figure*}

To highlight the dependence of excitation on the emission intensities, we plot the ratio of molecular emission over the column density ratio $I(X)/N(X)$ 
with respect to the lowest excitation case (Figure \ref{fig:intcol}). 
It is a simple measure to examine how modeled intensities are dependent on whether abundances or physical conditions (i.e., $n$ and $T$).
In Figure \ref{fig:intcol} ($left$), we compare the density variation of $I(X)/N(X)$ normalized to the $I(X)/N(X)$ in 
the case of $n=3\times 10^2\,$cm$^{-3}$ with other parameters being $A_{\rm}=2$\,mag, $\zeta = 10^{-17}\,$s$^{-1}$, $S/H=8\times 10^{-8}$, and $T=10\,$K.
Figure \ref{fig:intcol} ($right$) compares the variation on the temperature, and values of $I(X)/N(X)$ are normalized to the case of $T=10\,$K
with $A_{\rm}=2$\,mag, $\zeta = 10^{-17}\,$s$^{-1}$, $S/H=8\times 10^{-8}$, and $n=3\times 10^{3}\,$cm$^{-3}$.
Note that the above cases are only examples because the emission intensities are dependent on the optical depth, and dependence of intensity on the density and the temperature
may vary for other cases.
The dependence on the density is large for species other than CO and CI. On the other hand, ratios between the intensities of those species that we used for our analysis 
only vary by at most a factor of 3-4 when we vary $n = 3\times 10^2 - 3\times 10^4\,$cm$^{-3}$.
Similarly, the increase in the temperature causes higher intensities for all the species, but the variation between the species is kept within a factor of a few.

\begin{figure*}
\gridline{\fig{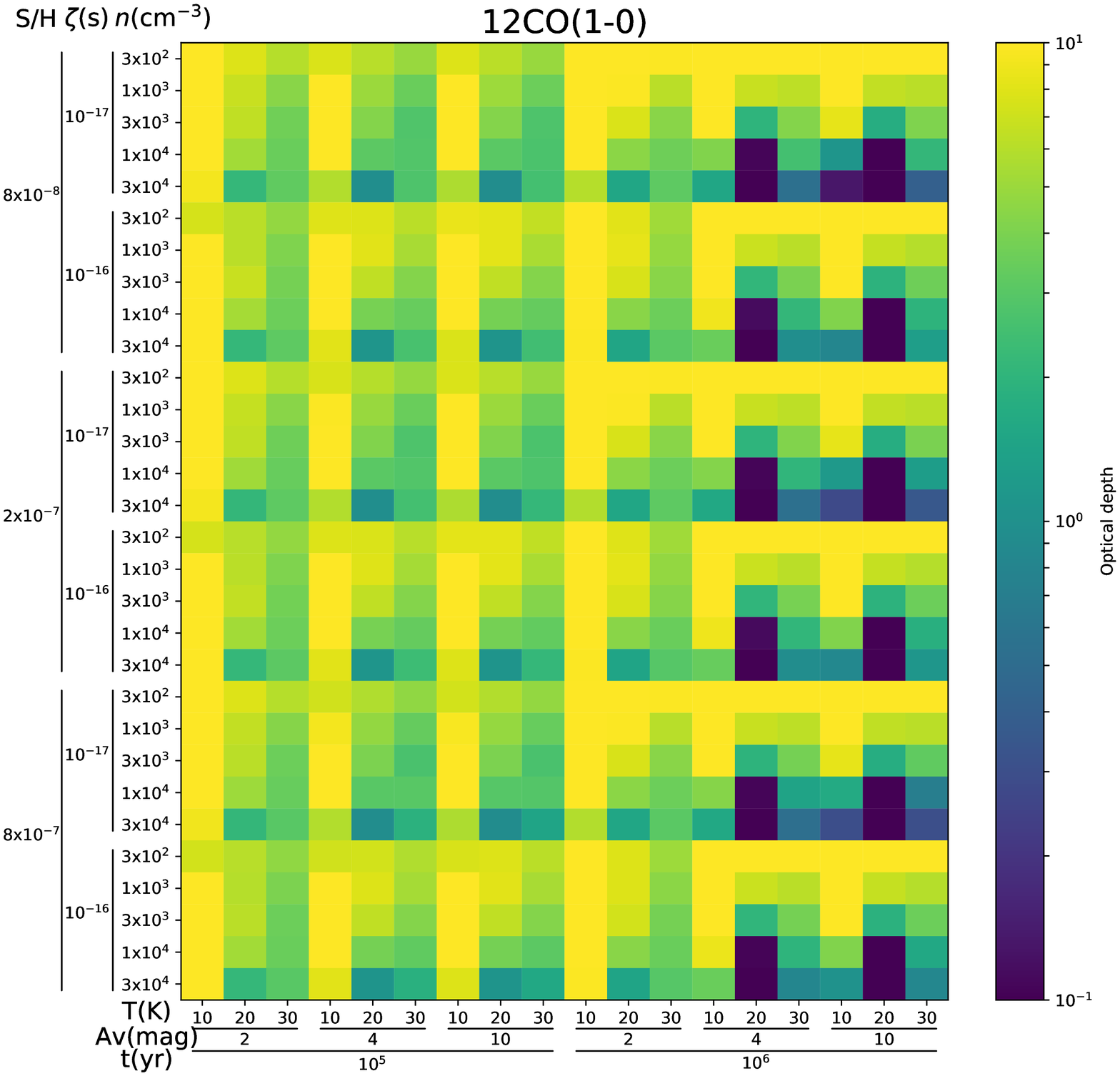}{0.50\textwidth}{  (a)}
\fig{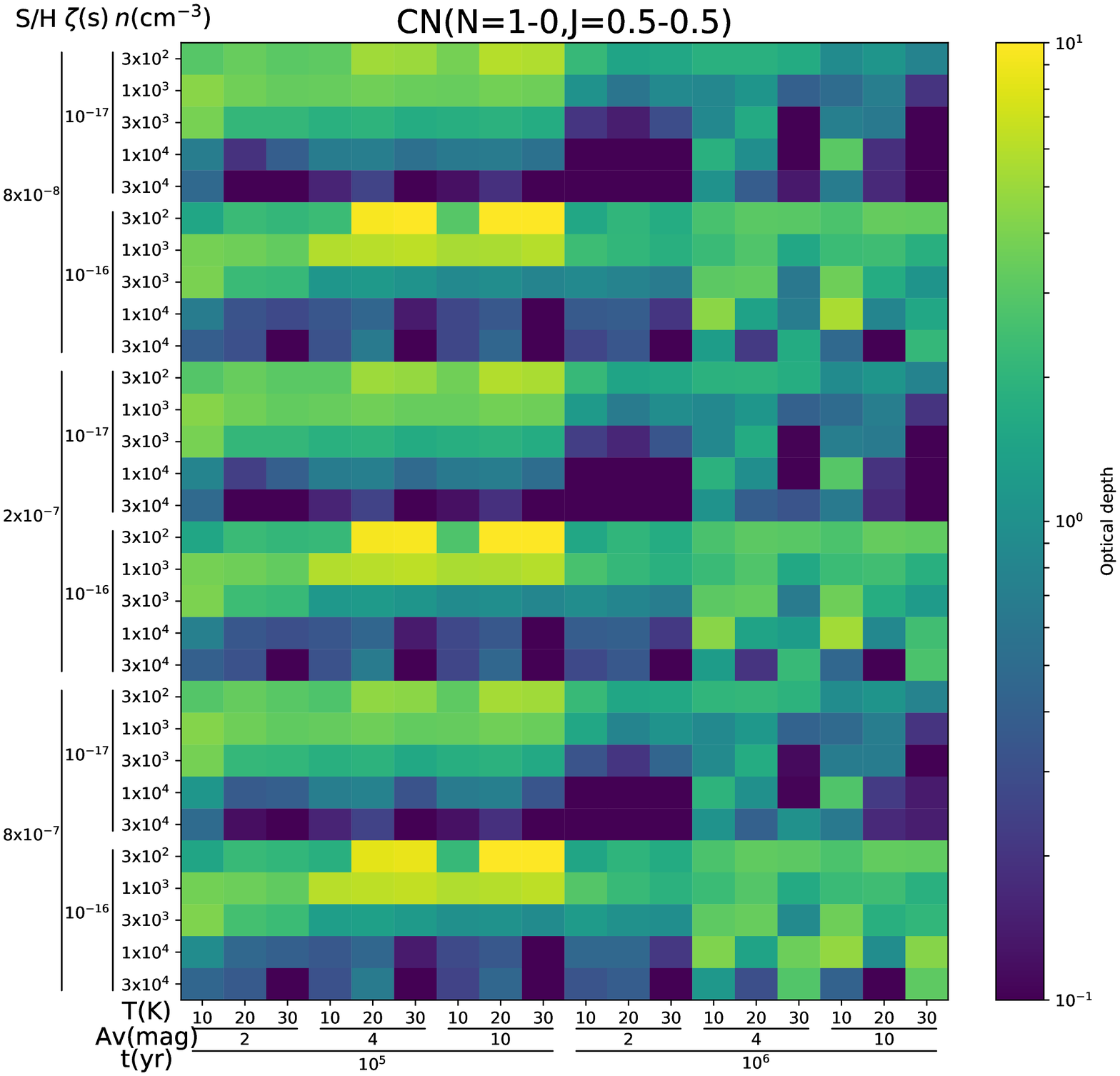}{0.5\textwidth}{(b)}
}

\gridline{\fig{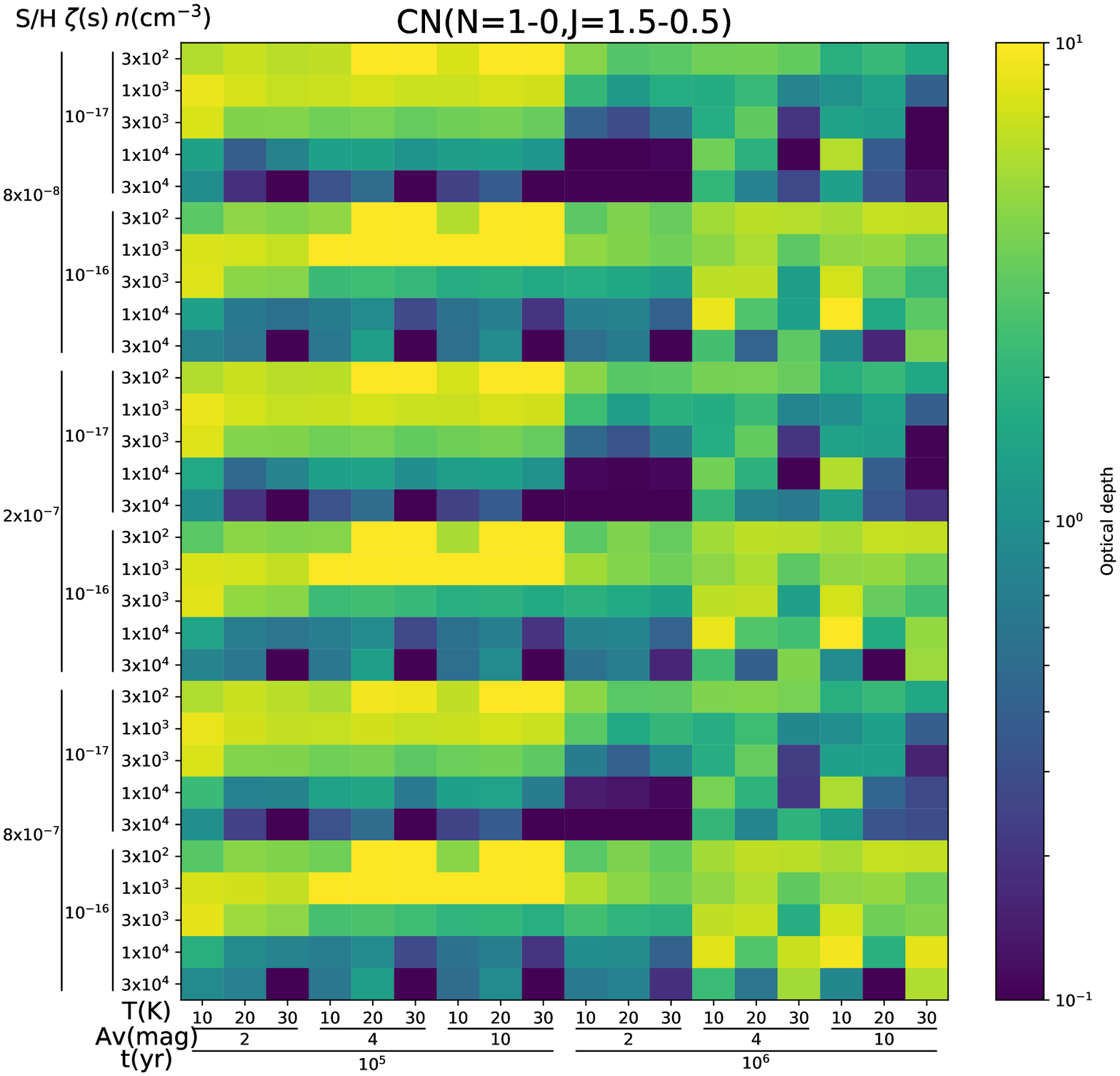}{0.5\textwidth}{ (c)}
\fig{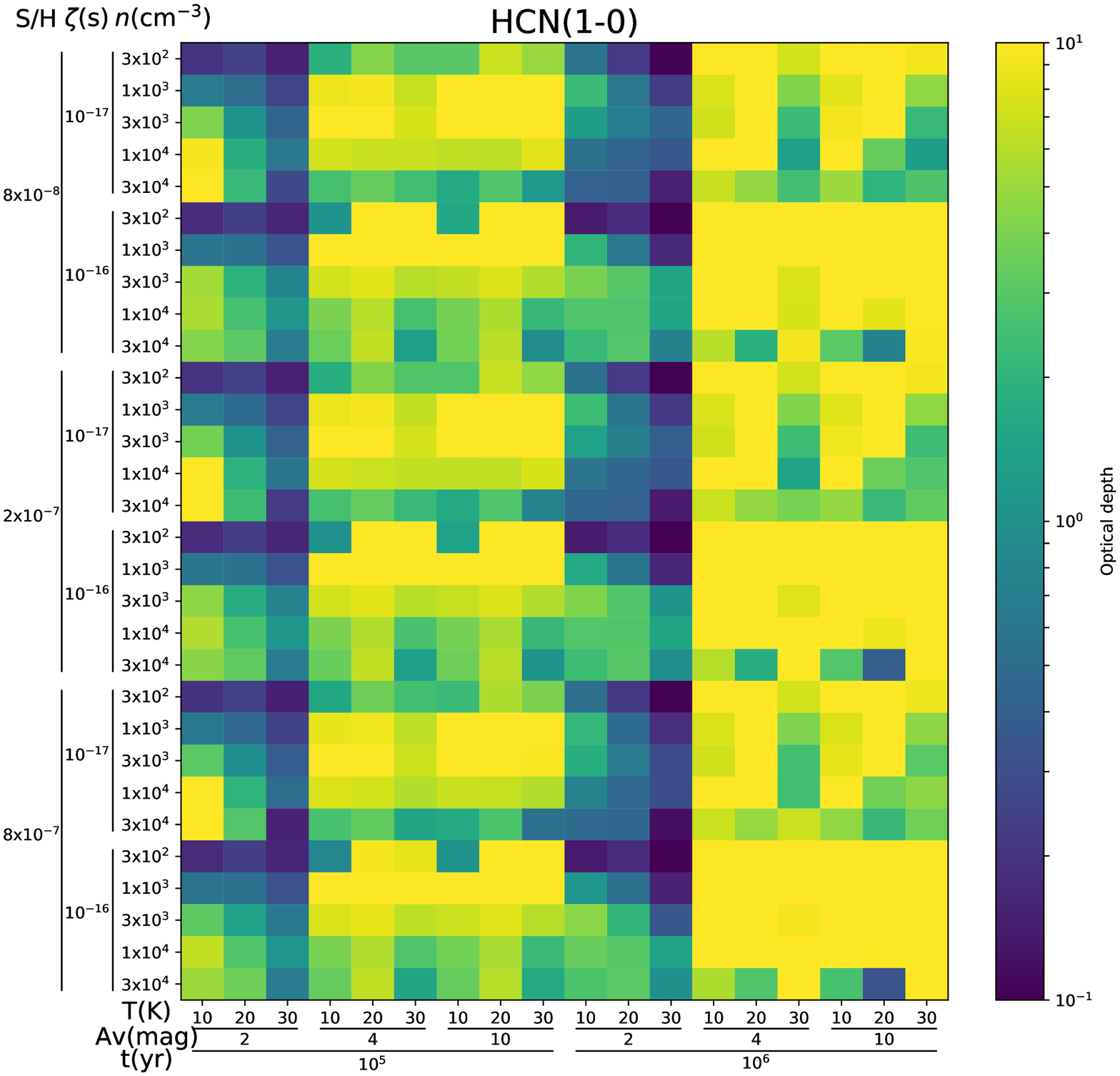}{0.5\textwidth}{(d)}
}

\caption{Optical depth at each model is shown for selected species. \label{fig:tau1}}
\end{figure*}

\begin{figure*}
\gridline{\fig{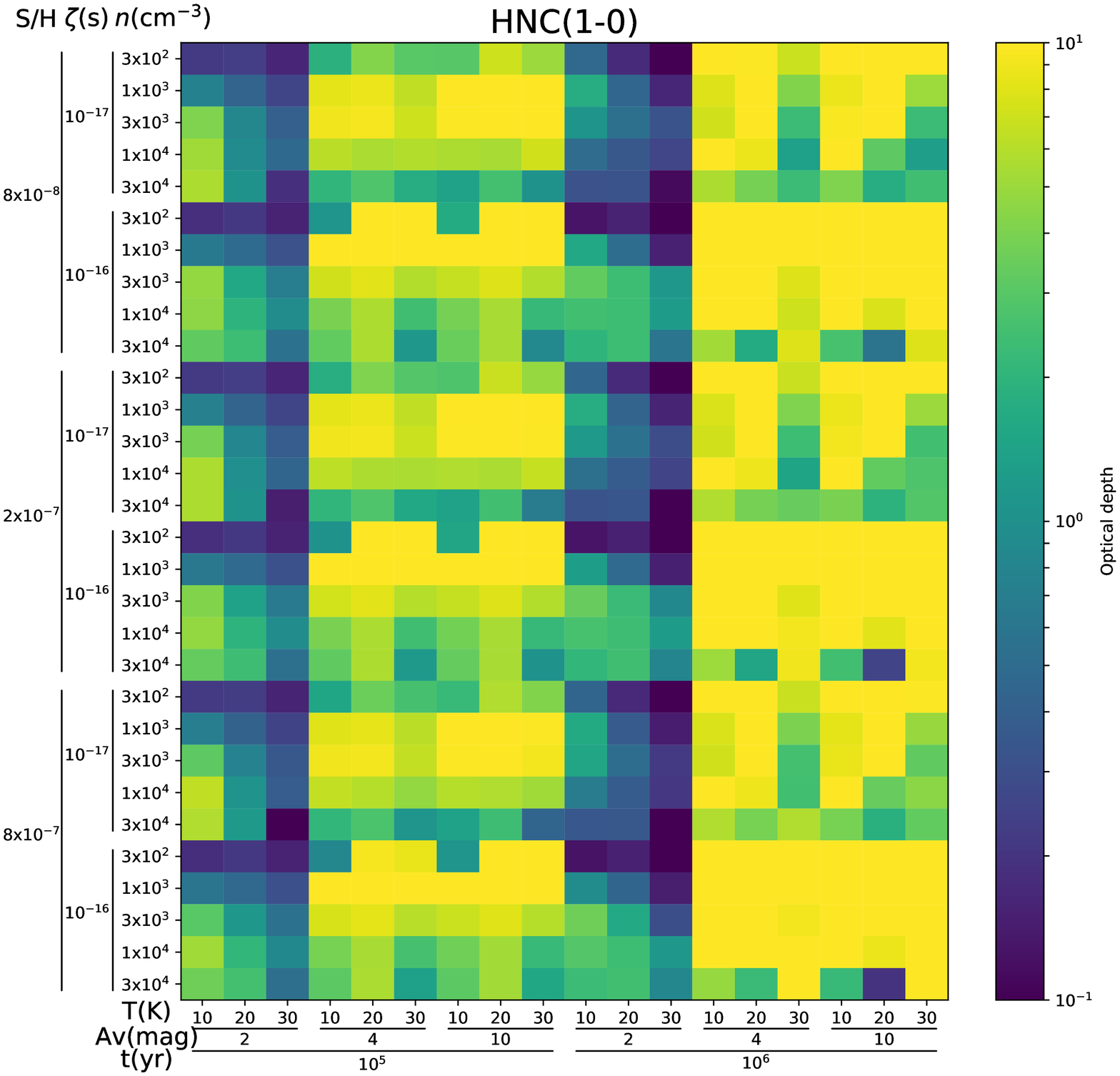}{0.50\textwidth}{ (a)}
\fig{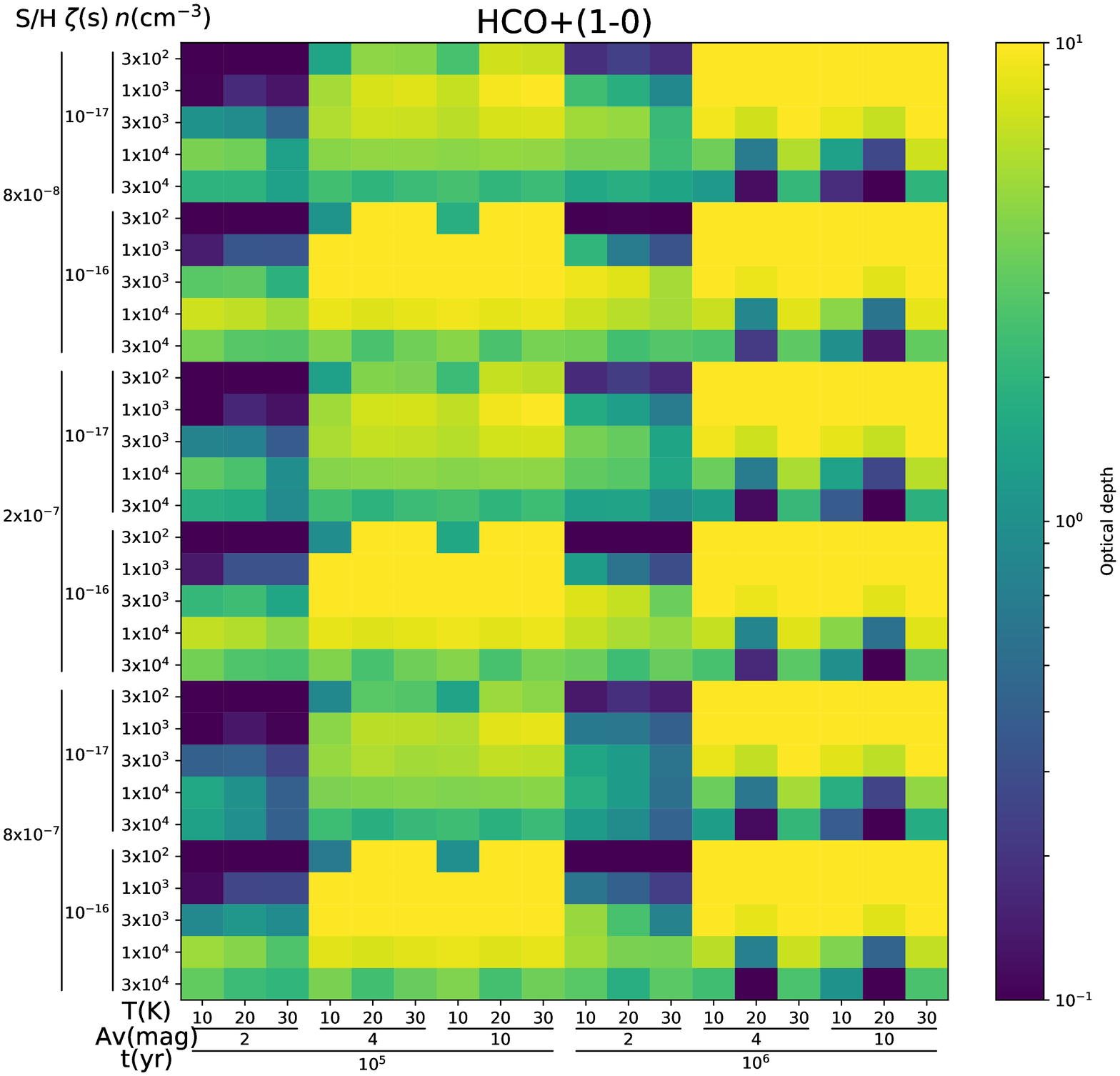}{0.5\textwidth}{(b)}
}

\gridline{\fig{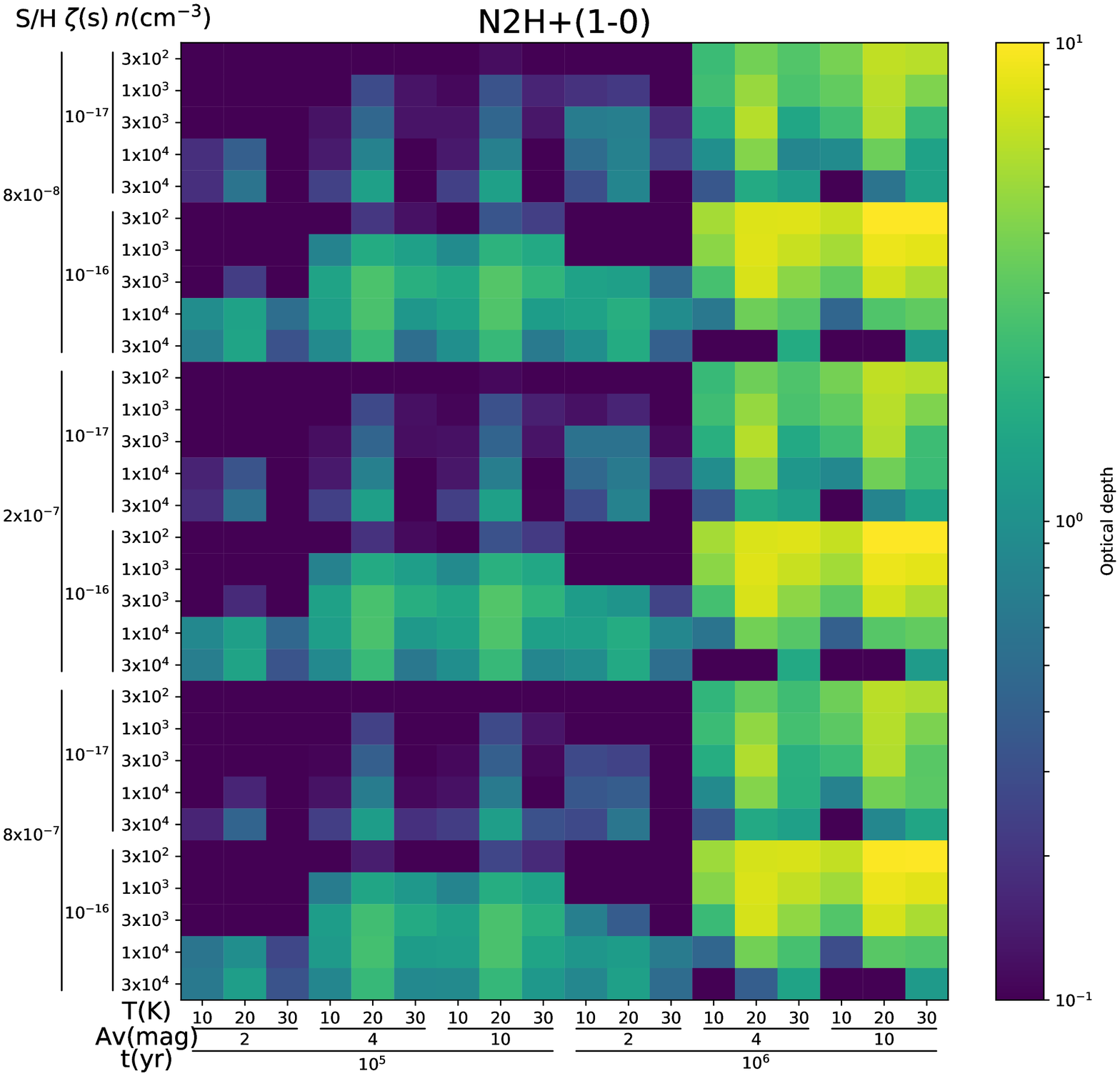}{0.5\textwidth}{(c)}
\fig{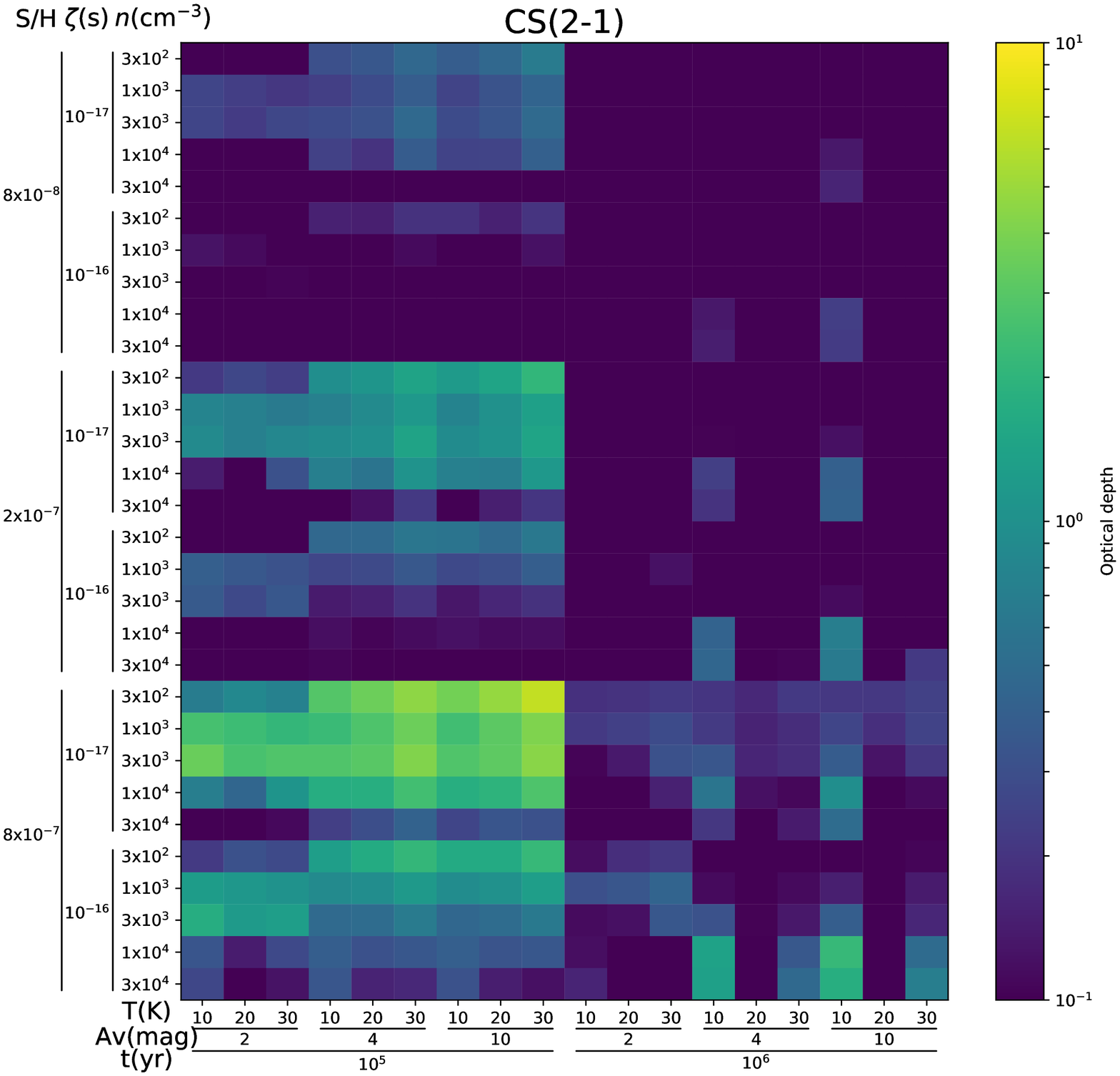}{0.5\textwidth}{(d)}
}

\caption{Same as Figure \ref{fig:tau1}, but for other species. \label{fig:tau2}}
\end{figure*}

\begin{figure*}
\gridline{\fig{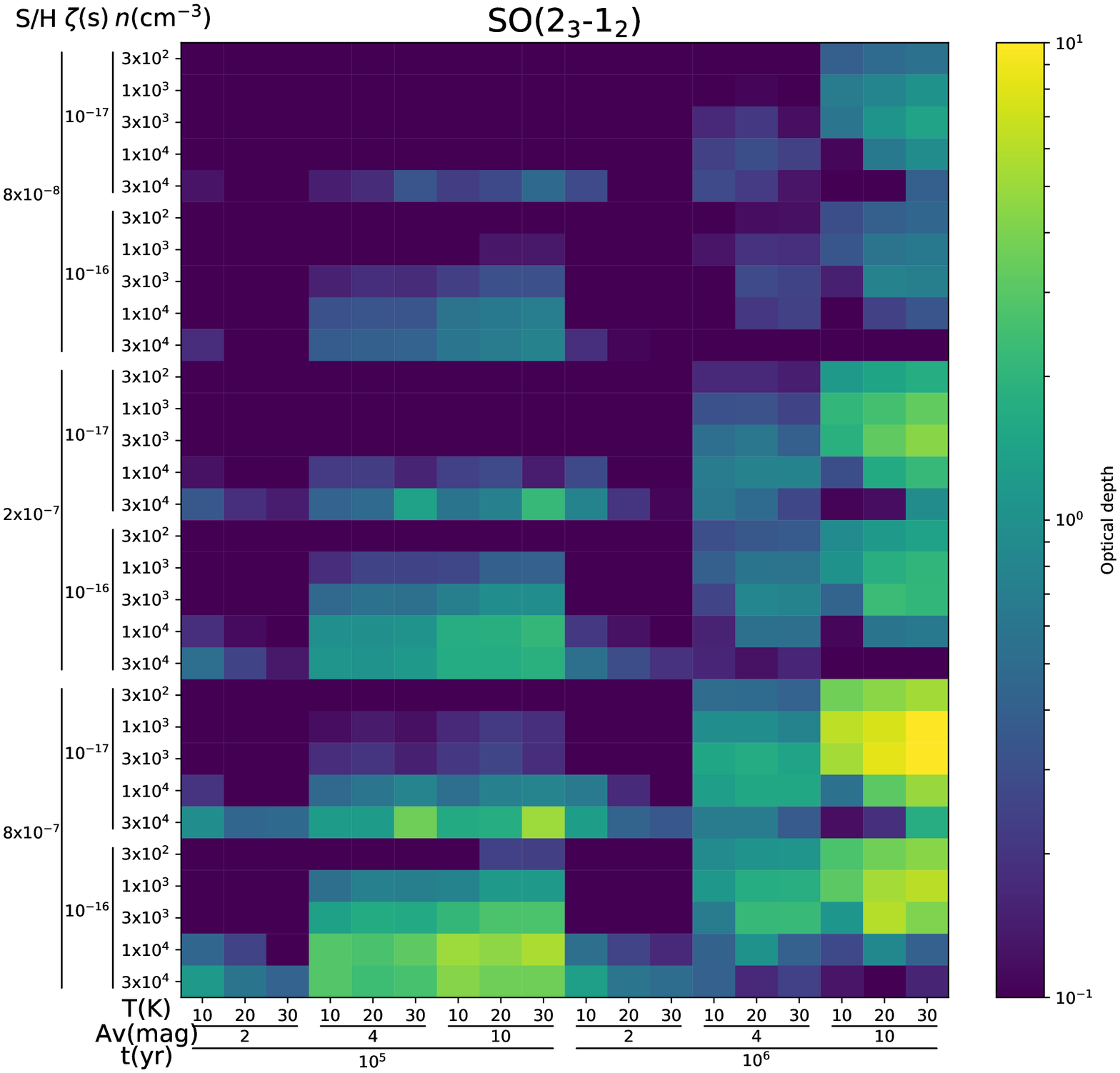}{0.50\textwidth}{(a)}
\fig{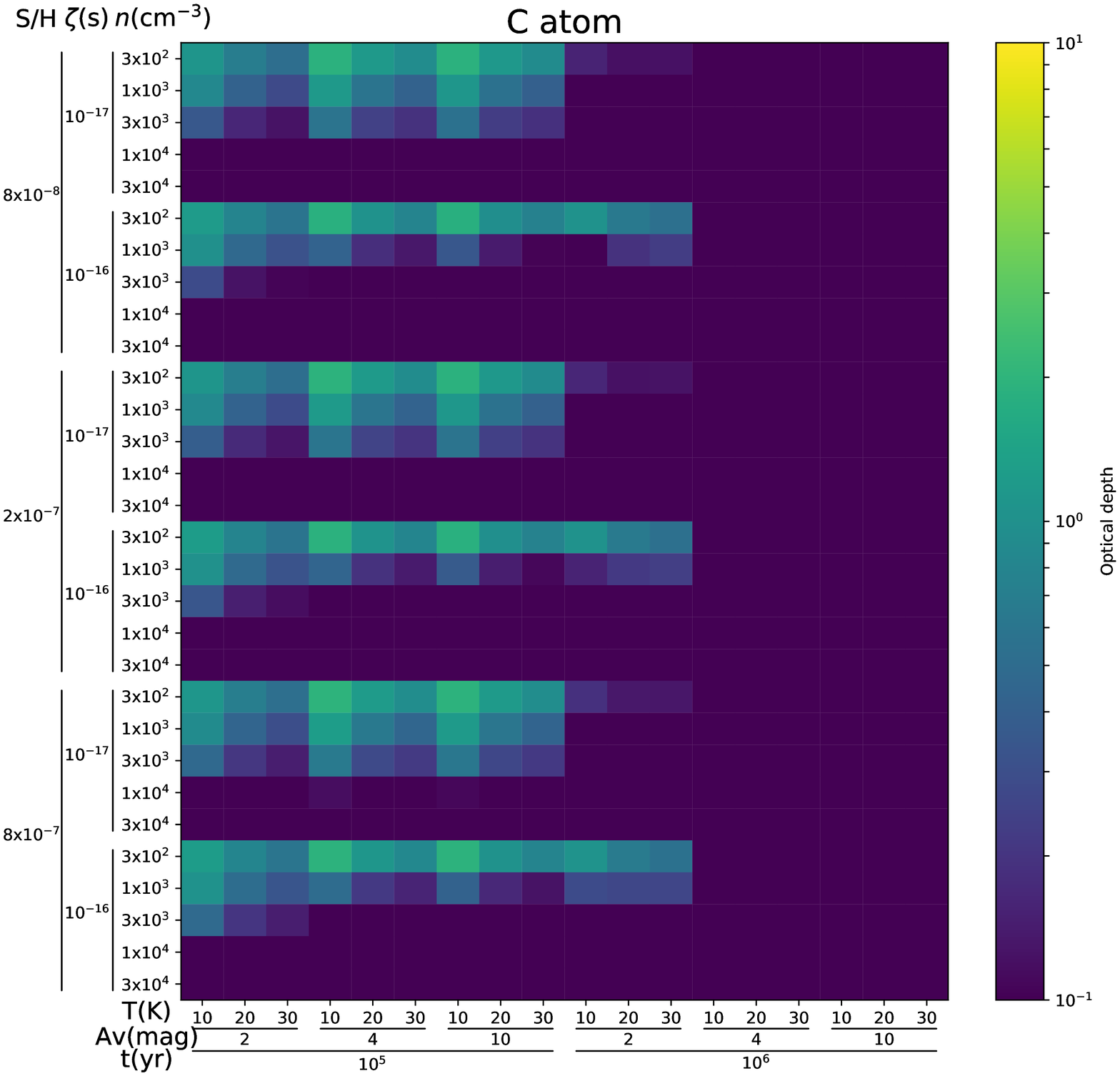}{0.5\textwidth}{ (b)}
}

\caption{Same as Figure \ref{fig:tau1}, but for other species. \label{fig:tau3}}
\end{figure*}

In our model calculations, there are transitions with large optical depth ($\tau \gg 1$) in some conditions 
while there are transitions that are optically thin in all the cases.
Figures \ref{fig:tau1} - \ref{fig:tau3} show the optical depth in our model calculations for transitions 
that have cases with $\tau \gtrsim 1$. Note that Figures \ref{fig:tau1} - \ref{fig:tau3} only use the color scale range of $10^{-1} - 10$
to show whether the condition is optically thin or thick.

\end{document}